\documentclass[a4paper,11pt]{article}
\pdfoutput=1

\usepackage{jheppub}
\usepackage{customprelude}

\title{Holography and discrete theta angles for disconnected gauge groups}

\author{Felix B. Christensen}

\affiliation{Department of Mathematical Sciences,
  Durham University,\\
  Durham, DH1 3LE, United Kingdom}
  
\emailAdd{felix.b.christensen@durham.ac.uk}

\abstract{Starting from holography, we derive the symmetry TFT for $\mathcal{N}=4$ SYM with $\mathfrak{so}(2n)$ gauge algebra, including terms which were previously unaccounted for holographically, by considering the action of the symmetry TFT on manifolds with torsion cycles. We then study gapped boundary conditions of the symmetry TFT and show how they correspond to the global forms and discrete theta angles studied by Hsin and Lam, including their higher group and non-invertible symmetries and anomalies. In particular, we analyse the case of theories with disconnected gauge groups. Considering the action of S-duality on the boundary conditions then leads to predictions for S-duality between these theories.}

\begin{document}
\maketitle
\flushbottom

\section{Introduction}
\label{sec:intro}
In this paper we consider the construction of IIB string theory on an $\mathcal{M}^5\times\mathbb{RP}^5$ orientifold, where $\mathcal{M}^5$ is taken to be asymptotically anti de Sitter. In \cite{Witten_1998} it was argued that this is holographically dual to $\mathcal{N}=4$ SYM with gauge algebra $\mathfrak{so}(n)$ and $\mathfrak{sp}(n)$. Our aim is to understand the possible global forms of the gauge theory in terms of boundary conditions of IIB supergravity in the $\mathfrak{so}(2N)$ case.

It was discovered in \cite{Gaiotto:2010be} that for gauge theories with non-simply connected gauge groups,  there are additional data needed to specify the global form of a theory and in particular the spectrum of line operators, called discrete theta angles. In \cite{Aharony:2013hda}, these were studied systematically for theories with connected gauge groups and related to certain topological terms in the action. Holographic accounts were given in \cite{Bergman_2022} and \cite{Etheredge:2023ler}. The analysis was generalised to disconnected gauge groups in \cite{Hsin_2021}. 

The aim of this paper is to give a holographic interpretation of the results of the latter paper. We first show how the symmetry TFT \cite{Gaiotto:2020iye,Apruzzi:2021nmk,Freed:2022qnc} and the topological theories on defect worldvolumes can be obtained from IIB supergravity and brane worldvolume couplings in the framework of differential cohomology. Similar analyses have been carried out in \cite{GarciaEtxebarria:2022vzq,Apruzzi:2022rei,Heckman:2022muc,Etheredge:2023ler} but we derive terms which have not previously been accounted for in this framework. In previous papers on this topic, $\mathcal{M}^5$ was assumed to be torsion free in cohomology. This meant that the final term of the K\"unneth short exact sequence given by 
\begin{equation}
	0\rightarrow\bigoplus_{i+j=n}H^i(C)\otimes H^j(C')\rightarrow H^n(C\otimes  C')\rightarrow\bigoplus_{i+j=n+1}\text{Tor}_1(H^i(C), H^j(C'))\rightarrow 0
\end{equation}
was trivial. We will see in section \ref{sec:couplings} that relaxing this assumption leads to additional terms both on the defect worldvolumes and the symmetry TFT action, which were expected from field theory considerations.

The rest of the paper is devoted to studying gapped boundary conditions of this action and matching them to global forms of $\mathfrak{so}(2N)$ gauge theory with disconnected gauge group and their generalised symmetries \cite{Gaiotto:2014kfa}. We shall see how the different boundary conditions reproduce the anomalies described in \cite{Hsin_2021} and identify their higher group \cite{Sharpe_2015,Tachikawa:2017gyf,Benini:2018reh,Bhardwaj:2021wif,DelZotto:2022joo,Apruzzi:2021vcu} and non-invertible \cite{Heidenreich:2021xpr,Kaidi:2021xfk,Choi:2021kmx} symmetries. In the final section we compute examples of fusion rules for these boundary conditions using the worldvolume actions of branes. Symmetries and discrete theta angles of gauge theories with orthogonal gauge groups in various dimensions have also been investigated in \cite{Cordova_2018,Bhardwaj_2023,Bartsch_2024,Bartsch_2024i,Bergman:2024its,Bonetti:2024etn}.

\subsection{Summary of results}
Throughout the paper, we consider only the cases of $\mathfrak{so}(8k)$ and $\mathfrak{so}(8k+2)$. Note first that the $\mathfrak{so}(8k+2)$ and $\mathfrak{so}(8k+6)$ theories differ only in the assignment of charges to line operators, while their symmetries and the form of their duality orbits are the same \cite{Aharony:2013hda}.\footnote{Moreover, we cannot yet account for the different charges from string theory, as explained in \cite{Etheredge:2023ler}.} For theories with connected gauge groups, there are different duality orbits for $\mathfrak{so}(8k)$ and $\mathfrak{so}(8k+4)$ theories. However, our main interest are theories with disconnected gauge groups, obtained by gauging the 0-form symmetry. It turns out that we can do this only for duality orbits which are the same for $\mathfrak{so}(8k)$ and $\mathfrak{so}(8k+4)$. All other orbits suffer from a gauge-global anomaly preventing us from gauging the 0-form symmetry. It is therefore sufficient to consider only these two cases instead of all four. The symmetry TFT is given by 
\begin{equation}
	S_{5D}=i\pi\int_{\mathcal{M}^5}a_1\smile \delta a_3+b_{F1}\smile\delta b_{NS5}+c_{D1}\smile\delta c_{D5}+a_1\smile b_{F1}\smile c_{D1}+\frac{N}{2}\delta b_{F1}\smile c_{D1}
\end{equation}
where all of the fields are $\mathbb{Z}_2$ gauge fields. $a_1$ couples electrically to D3 branes wrapping $\mathbb{RP}^3\subset\mathbb{RP}^5$, and $a_3$ couples electrically to D3 branes wrapping $\mathbb{RP}^1\subset\mathbb{RP}^5$. All other fields are 2-form gauge fields which couple electrically to the objects specified in the subscript. Here, D5 and NS5 refers to ``fat strings", i.e. branes wrapping a twisted cocycle represented by $\mathbb{RP}^4\subset\mathbb{RP}^5$ \cite{Witten_1998}. We are mainly interested in the cases where $a_3$ has Dirichlet boundary conditions so we will not write this down explicitly for each global form.
\subsubsection{$\mathfrak{so}(8k)$}
One set of duality orbits is summarised in figure \ref{figure:so8k1}. Here, and in subsequent figures, the top line describes the global form of the gauge group and the following lines specify which combinations of gauge fields have Dirichlet boundary conditions:

\begin{figure}
	\centering
	\begin{tikzcd}		
		\left(
		\begin{tabular}{p{3cm}}
			Pin$^+(8k)$\\	
			$b_{F1}$\\ 
			$c_{D5}$\\
		\end{tabular}
		\right)\arrow[loop left, looseness=3]{}{\text{T}}\arrow[d,"\text{S}"]&\hspace{0.5cm}\left(\begin{tabular}{p{3cm}}
			O$(8k)_{-+}$\\	
			$b_{F1}$\\ 
			$c_{D5}+c_{D1}$\\
		\end{tabular}\right)\arrow[loop right, looseness=3]{}{\text{T}}\arrow[d,"\text{S}"]\\
		\left(
		\begin{tabular}{p{3cm}}
			PO$(8k)_{\begin{smallmatrix}+&+&+\\+&+&\end{smallmatrix}}$\\	
			$b_{NS5}$\\ 
			$c_{D1}$\\
		\end{tabular}
		\right)\arrow[u]\arrow[d,"\text{T}"]&\hspace{0.5cm}\left(\begin{tabular}{p{3cm}}
			PO$(8k)_{\begin{smallmatrix}-&-&+\\-&-&\end{smallmatrix}}$\\	
			$c_{D1}$\\ 
			$b_{NS5}+b_{F1}$\\
		\end{tabular}\right)\arrow[u]\arrow[d,"\text{T}"]\\
		\left(
		\begin{tabular}{p{3cm}}
			PO$(8k)_{\begin{smallmatrix}-&+&+\\+&-&\end{smallmatrix}}$\\	
			$b_{F1}+c_{D1}$\\ 
			$c_{D5}+b_{NS5}$\\
		\end{tabular}
		\right)\arrow[loop left, looseness=3]{}{\text{S}}\arrow[u]&\hspace{0.5cm}\left(\begin{tabular}{p{3cm}}
			PO$(8k)_{\begin{smallmatrix}+&-&+\\-&+&\end{smallmatrix}}$\\	
			$b_{F1}+c_{D1}$\\ 
			$b_{F1}+b_{NS5}+c_{D5}$\\
		\end{tabular}\right)\arrow[loop right, looseness=3]{}{\text{S}}\arrow[u]\\
	\end{tikzcd}
	\begin{tikzcd}
			\left(\begin{tabular}{p{3cm}}
				O$(8k)_{++}$\\	
				$b_{F1}$\\ 
				$c_{D1}$\\
			\end{tabular}\right)\arrow[loop left, looseness=3]{}{\text{S}}\arrow[loop right, looseness=3]{}{\text{T}}
	\end{tikzcd}
	\caption{Simple $\mathfrak{so}(8k)$ duality orbits}\label{figure:so8k1}
\end{figure}
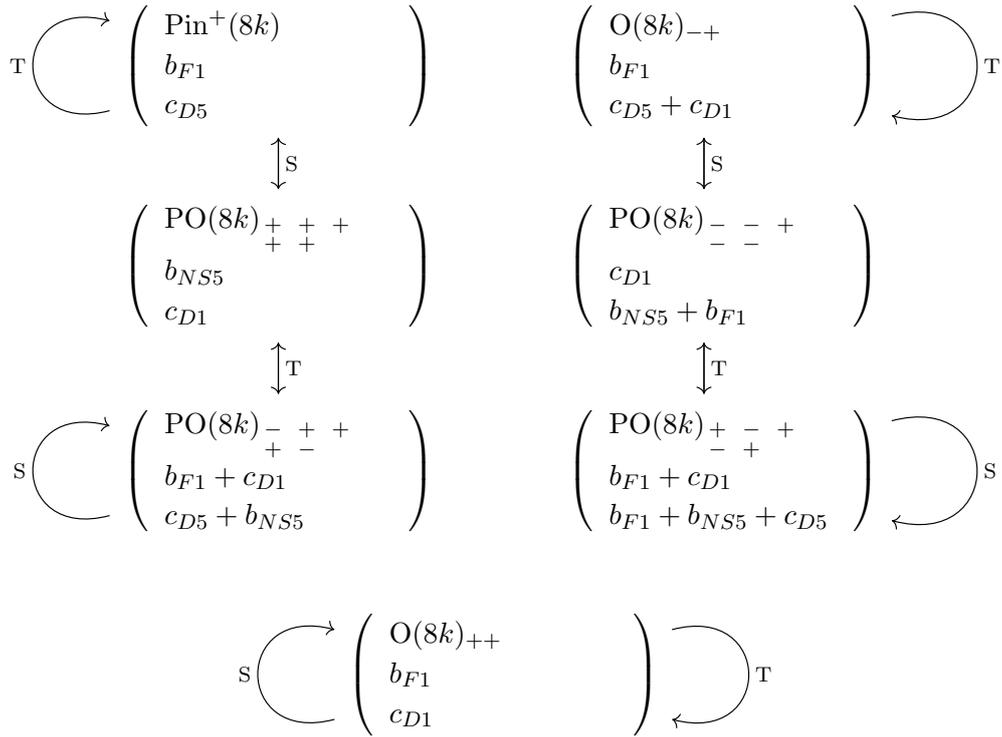
The singlet has a 3-group symmetry while in the other two duality orbits, the 1-form symmetry becomes non-invertible. Note that the duality orbits simply mirror the ones shared between $\mathfrak{so}(8k)$ and $\mathfrak{so}(8k+4)$ theories found in \cite{Aharony:2013hda}.

As we shall see, we are also required to consider boundary conditions mixing gauge fields of different degrees. We will call these boundary conditions ``non-simple". The resulting orbits are summarised in figure \ref{figure:so8k2}.
\begin{figure}
	\centering
	\begin{tikzcd}
	\left(
	\begin{tabular}{p{4cm}}
		$\widetilde{\text{Pin}}(8k)$\\	
		$b_{F1}$\\ 
		$c_{D5}-(a_1)^2$\\
	\end{tabular}
	\right)\arrow[loop left, looseness=3]{}{\text{T}}\arrow[d,"\text{S}"]&\hspace{-0.5cm}\left(\begin{tabular}{p{4cm}}
		O$(8k)_{--}$\\	
		$b_{F1}$\\ 
		$c_{D5}+c_{D1}-(a_1)^2$\\
	\end{tabular}\right)\arrow[loop right, looseness=3]{}{\text{T}}\arrow[d,"\text{S}"]\\
	\left(
	\begin{tabular}{p{4cm}}
		PO$(8k)_{\begin{smallmatrix}+&+&-\\+&+&\end{smallmatrix}}$\\	
		$c_{D1}$\\
		$b_{NS5}-(a_1)^2$\\ 
	\end{tabular}
	\right)\arrow[u]\arrow[d,"\text{T}"]&\hspace{-0.5cm}\left(\begin{tabular}{p{4cm}}
		PO$(8k)_{\begin{smallmatrix}-&-&-\\-&-&\end{smallmatrix}}$\\	
		$c_{D1}$\\ 
		$b_{NS5}+b_{F1}-(a_1)^2$\\
	\end{tabular}\right)\arrow[u]\arrow[d,"\text{T}"]\\
	\left(
	\begin{tabular}{p{4cm}}
		PO$(8k)_{\begin{smallmatrix}-&+&-\\+&-&\end{smallmatrix}}$\\	
		$b_{F1}+c_{D1}$\\ 
		$c_{D5}+b_{NS5}-(a_1)^2$\\
	\end{tabular}
	\right)\arrow[loop left, looseness=3]{}{\text{S}}\arrow[u]&\hspace{-0.5cm}\left(\begin{tabular}{p{4cm}}
		PO$(8k)_{\begin{smallmatrix}+&-&-\\-&+&\end{smallmatrix}}$\\	
		$b_{F1}+c_{D1}$\\ 
		$b_{F1}+b_{NS5}+c_{D5}-(a_1)^2$\\
	\end{tabular}\right)\arrow[loop right, looseness=3]{}{\text{S}}\arrow[u]\\
\end{tikzcd}
\begin{tikzcd}
	\left(\begin{tabular}{p{1.9cm}}
		O$(8k)_{+-}$\\	
		$b_{F1}$\\ 
		$c_{D1}-(a_1)^2$\\
	\end{tabular}\right)\arrow[loop left, looseness=3]{}{\text{T}}\arrow[r,"\text{S}"]&
	\left(\begin{tabular}{p{1.9cm}}
		Pin$^-(8k)_{+}$\\	
		$b_{F1}-(a_1)^2$\\ 
		$c_{D1}$\\
	\end{tabular}\right)\arrow[r,"\text{T}"]\arrow[l]&
	\left(\begin{tabular}{p{1.9cm}}
		Pin$^-(8k)_{-}$\\	
		$b_{F1}-(a_1)^2$\\ 
		$c_{D1}-(a_1)^2$\\
	\end{tabular}\right)\arrow[loop right, looseness=3]{}{\text{S}}\arrow[l]\\
\end{tikzcd}
\caption{Non-simple $\mathfrak{so}(8k)$ duality orbits}\label{figure:so8k2}
\end{figure}
The meaning of $\widetilde{\text{Pin}}(8k)$ will be explained later. We shall see that all of these duality orbits have non-invertible 1-form symmetries.
\subsubsection{$\mathfrak{so}(8k+2)$}
In the $\mathfrak{so}(8k+2)$ case, there is an additional subtlety, since for some global forms, the bulk $\mathbb{Z}_2$ gauge fields combine into $\mathbb{Z}_4$ background fields on the boundary. In such cases, we will denote the $\mathbb{Z}_4$ uplift of a gauge field $b$ by $\tilde{b}$. For our simple boundary conditions, we again find symmetry orbits mirroring the case with connected gauge groups as shown in figure \ref{figure:so8kplus21}.
\begin{figure}
	\centering
	\begin{tikzcd}
		&\left(\begin{tabular}{p{3cm}}
			Pin$^+(8k+2)$\\	
			$2\tilde{c}_{D5}+\tilde{b}_{F1}$\\ 
		\end{tabular}\right)\arrow[loop left, looseness=3]{}{\text{T}}\arrow[d,"\text{S}"]&\\
		&\left(\begin{tabular}{p{3cm}}
			PO$(8k+2)_{0+}$\\	
			$2\tilde{b}_{NS5}+\tilde{c}_{D1}$\\ 
		\end{tabular}\right)\arrow[dl,"\text{T}"]\arrow[u]&\\
		\left(\begin{tabular}{p{3cm}}
			PO$(8k+2)_{3+}$\\	
			$2(\tilde{b}_{NS5}+\tilde{c}_{D5})+\tilde{b}_{F1}+\tilde{c}_{D1}$\\ 
		\end{tabular}\right)\arrow[dr,"\text{T}"]\arrow[rr,"\text{S}"]&&\left(\begin{tabular}{p{3cm}}
		PO$(8k+2)_{1+}$\\	
		$2(\tilde{b}_{NS5}+\tilde{c}_{D5})+\tilde{b}_{F1}-\tilde{c}_{D1}$\\ 
		\end{tabular}\right)\arrow[ul,"\text{T}"]\arrow[ll]\\
		&\left(\begin{tabular}{p{3cm}}
			PO$(8k+2)_{2+}$\\	
			$2(\tilde{c}_{D5}+\tilde{b}_{F1})+\tilde{c}_{D1}$\\ 
		\end{tabular}\right)\arrow[d,"\text{S}"]\arrow[ur,"\text{T}"]&\\
		&\left(\begin{tabular}{p{3cm}}
			O$(8k+2)_{-+}$\\	
			$2(\tilde{c}_{D5}+\tilde{c}_{D1})+\tilde{b}_{F1}$\\ 
		\end{tabular}\right)\arrow[loop left, looseness=3]{}{\text{T}}\arrow[u]&\\
	\end{tikzcd}
	\begin{tikzcd}
		&\left(\begin{tabular}{p{3cm}}
			O$(8k+2)_{++}$\\	
			$c_{D1}$\\
			$b_{F1}$\\
		\end{tabular}\right)\arrow[loop left, looseness=3]{}{\text{T}}\arrow[loop right, looseness=3]{}{\text{S}}&\\
	\end{tikzcd}
	\caption{Simple $\mathfrak{so}(8k+2)$ duality orbits}\label{figure:so8kplus21}
\end{figure}
The duality orbits for non-simple boundary conditions are summarised in figure \ref{figure:so8kplus22}.
\begin{figure}
	\centering
	\begin{tikzcd}
		&\left(\begin{tabular}{p{3.4cm}}
			$\widetilde{\text{Pin}}(8k+2)$\\	
			$2(\tilde{c}_{D5}-(\tilde{a}_1)^2)+\tilde{b}_{F1}$\\ 
		\end{tabular}\right)\arrow[loop left, looseness=3]{}{\text{T}}\arrow[d,"\text{S}"]&\\
		&\left(\begin{tabular}{p{3.4cm}}
			PO$(8k+2)_{0-}$\\	
			$2(\tilde{b}_{NS5}-(\tilde{a}_1)^2)+\tilde{c}_{D1}$\\ 
		\end{tabular}\right)\arrow[dl,"\text{T}"]\arrow[u]&\\
		\left(\begin{tabular}{p{3cm}}
			PO$(8k+2)_{3-}$\\	
			$2(\tilde{b}_{NS5}+\tilde{c}_{D5}-(\tilde{a}_1)^2)+\tilde{b}_{F1}+\tilde{c}_{D1}$\\ 
		\end{tabular}\right)\arrow[dr,"\text{T}"]\arrow[rr,"\text{S}"]&&\left(\begin{tabular}{p{3cm}}
			PO$(8k+2)_{1-}$\\	
			$2(\tilde{b}_{NS5}+\tilde{c}_{D5}-(\tilde{a}_1)^2)+\tilde{b}_{F1}-\tilde{c}_{D1}$\\ 
		\end{tabular}\right)\arrow[ul,"\text{T}"]\arrow[ll]\\
		&\left(\begin{tabular}{p{3.4cm}}
			PO$(8k+2)_{2-}$\\	
			$2(\tilde{c}_{D5}+\tilde{b}_{F1}-(\tilde{a}_1)^2)+\tilde{c}_{D1}$\\ 
		\end{tabular}\right)\arrow[d,"\text{S}"]\arrow[ur,"\text{T}"]&\\
		&\left(\begin{tabular}{p{3.5cm}}
			O$(8k+2)_{--}$\\	
			$2(\tilde{c}_{D5}+\tilde{c}_{D1}-(\tilde{a}_1)^2)+\tilde{b}_{F1}$\\ 
		\end{tabular}\right)\arrow[loop left, looseness=3]{}{\text{T}}\arrow[u]&\\
	\end{tikzcd}
	\begin{tikzcd}
		\left(\begin{tabular}{p{2.1cm}}
			O$(8k+2)_{+-}$\\	
			$b_{F1}$\\ 
			$c_{D1}-(a_1)^2$\\
		\end{tabular}\right)\arrow[loop left, looseness=3]{}{\text{T}}\arrow[r,"\text{S}"]&
		\left(\begin{tabular}{p{2.4cm}}
			Pin$^-(8k+2)_{+}$\\	
			$b_{F1}-(a_1)^2$\\ 
			$c_{D1}$\\
		\end{tabular}\right)\arrow[r,"\text{T}"]\arrow[l]&
		\left(\begin{tabular}{p{2.4cm}}
			Pin$^-(8k+2)_{-}$\\	
			$b_{F1}-(a_1)^2$\\ 
			$c_{D1}-(a_1)^2$\\
		\end{tabular}\right)\arrow[loop right, looseness=3]{}{\text{S}}\arrow[l]\\
	\end{tikzcd}
	\caption{Non-simple $\mathfrak{so}(8k+2)$ duality orbits}\label{figure:so8kplus22}
\end{figure}
As before, the singlet has a 3-group symmetry while all other duality orbits have a non-invertible 1-form symmetry.
\section{SymTFT  couplings}\label{sec:couplings}
The symmetry TFT of the theories we are interested in contains both $BF$ terms and anomaly terms. The $BF$ terms can be derived from the requirement that the symmetry TFT operators match the non-commutativity of branes in \cite{GarciaEtxebarria:2022vzq}. We expect that one could also obtain them from the supergravity action as done in \cite{GarciaEtxebarria:2024fuk}. In this section we shall focus on a subtlety in deriving the anomaly terms.

It is well-known that the RR fields in IIB theory should be described by an (as yet unknown) S-duality covariant generalisation of twisted differential K-theory \cite{Moore:1999gb,Freed:2000ta,diaconescu2004e8gaugetheoryderivation,Evslin:2006cj}. However, for the purposes of this paper, we shall model them by ordinary differential cohomology. This will turn out to be sufficient to reproduce known results from the field theory literature, with one possible exception.  
\subsection{Differential cohomology and supergravity fields}
Since we are interested in understanding topological features of supergravity fields, it is convenient to use the language of differential cohomology \cite{10.1007/BFb0075216}. For a comprehensive mathematical review, see \cite{bär2014differential}, while more detailed accounts aimed at physicists can be found in \cite{Freed:2006yc,Apruzzi:2021nmk}. Here, we briefly recall some facts which will be important later. 

A differential character on a manifold $\mathcal{M}$ is an element $\chi\in\text{Hom}(Z_{\ell-1},U(1))$ such that there exists $F_{\chi}\in\Omega^\ell(\mathcal{M})$ with $\chi(\Sigma)=\exp(2\pi i\int_{\mathcal{B}}F_{\chi})$ for all $\Sigma$, $\mathcal{B}$ such that $\Sigma=\partial\mathcal{B}$. The resulting groups of differential characters are denoted $\breve{H}^{\ell}(\mathcal{M})$ and are called differential cohomology groups. If $\mathcal{M}$ is a $d$-dimensional manifold, the differential cohomology groups with $0\le\ell\le d+1$ may be nontrivial. One benefit of differential cohomology is that it contains precisely the gauge-invariant information of a (higher) gauge field. For example, it can be shown that $\breve{H}^2(\mathcal{M})$ is isomorphic to the group of isomorphism classes of $U(1)$ bundles with connection over $\mathcal{M}$, and $F_{\chi}$ is exactly the field strength. As mentioned before, in the case of string theory, the RR gauge fields are actually described by a generalised differential cohomology theory, however we shall see that approximating it by ordinary differential cohomology is sufficient for our purposes.

To discuss further useful features of differential cohomology, we consider a model of differential cohomology developed by Hopkins and Singer \cite{Hopkins:2002rd}. This has been reviewed in the physics literature in \cite{Belov:2006jd,Belov:2006xj,Hsieh:2020jpj,GarciaEtxebarria:2024fuk}. A Hopkins-Singer cocycle consists of the data $(c,h,\omega)\in Z^n(\mathcal{M};\mathbb{Z})\times C^{n-1}(\mathcal{M};\mathbb{R})\times \Omega^n(\mathcal{M})$ with $d\omega=0$, $\delta c=0$, and $\delta h =-c+\omega$ where inclusion maps are implied. The cohomology class $[c]$ coincides with the Chern class of the corresponding $U(1)$ bundle in the case of $\breve{H}^2(\mathcal{M})$ and in general, it is called the characteristic class of a differential character. $h$ is related to the differential character $\chi$ represented by a cocycle $(c,h,\omega)$. For every cycle $\Sigma\in Z^{n-1}(\mathcal{M})$, we have
\begin{equation}
	\chi(\Sigma)=\exp(2\pi i h(\Sigma))\, .
\end{equation}
$\omega$ is known as the curvature of the differential cocycle. Physically, it corresponds to a field strength.

One can verify that the holonomy is left invariant by gauge transformations of the form $(c,h,\omega)\rightarrow(c+\delta\lambda,h+\lambda+\delta r,\omega)$ where $\lambda\in C^{n-1}(\mathcal{M};\mathbb{Z})$ is the gauge parameter for large gauge transformations while $r\in C^{n-2}(\mathcal{M};\mathbb{R})$ corresponds to small gauge transformations. Taking the quotient of Hopkins-Singer cocycles by such gauge transformations results in the differential cohomology group $\breve{H}^n(\mathcal{M})$. We omit a discussion of how this fits into a double cochain complex.

We may define a product on differential cohomology, denoted by $\star$. At the level of Hopkins-Singer cocycles it is given by $(c,h,\omega)\star(c',h',\omega')=(c\smile c',(-1)^{|c|}c\smile h' + h\smile \omega' + B(\omega,\omega'),\omega\wedge\omega')$. Here, $B$ is any natural chain homotopy between the wedge product and the cup product, such as the one explicitly constructed in \cite{10.1007/BFb0075216}. This product has the Leibniz property and therefore passes to a well-defined product on differential characters. It also turns out to be graded commutative on differential characters, however showing this is rather involved.
 Finally, we will need the notion of fibre-integration. For the purposes of this paper, we can restrict ourselves to trivial fibrations $\mathcal{M}\times \Phi$ for $\Phi$ $p$-dimensional and oriented. In this case, fibre integration is a map given by 
 
 \begin{equation}
 	\int_{\Phi}(c,h,\omega)=(c/\Phi,h/\Phi,\omega/\Phi)
 \end{equation}
where $\Phi\in Z_n(\Phi)$ is a representative of the fundamental class of $\Phi$ by a slight abuse of notation. The slant product \cite{Hatcher} is defined as follows: let $c\in C^l(\mathcal{M}\times \Phi,A)$ and $v\in C_p(\Phi,A)$. Then $c/v\in C^{l-p}(\mathcal{M},A)$ is a cochain such that for all $u\in C_{l-p}(\mathcal{M})$ the equation $c/v(u)=c(v\times u)$ holds. It can be shown that this descends to a well-defined integration map
\begin{equation}
	\int_{\Phi}: \breve{H}^p(\mathcal{M}\times\Phi)\rightarrow\breve{H}^{p-\text{dim}(\Phi)}(\mathcal{M})\, .
\end{equation}
\subsection{Dimensional reduction}
In this section, we will see how carefully performing the dimensional reduction while allowing for torsion cycles on our spacetime $\mathcal{M}^5$ affects our ansatz for supergravity fields in the topological sector. This will later lead to additional couplings in the symmetry TFT and on the worldvolumes of symmetry defects.

Compactification in differential cohomology is discussed in \cite{Apruzzi:2021nmk} and the discussion is extended to the Hopkins-Singer model in \cite{GarciaEtxebarria:2024fuk}.
In both cases, it is assumed that the spacetime is of the form $\mathcal{M}^p\times X^q$ where $\mathcal{M}^p$ is taken to be torsion free. In this case, the third term in the Künneth short exact sequence is trivial and it follows that $H^*(\mathcal{M}^p\times X^q;\mathbb{Z})\cong H^*(\mathcal{M}^p;\mathbb{Z}) \otimes H^*(X^q;\mathbb{Z})$. However, from the above discussion we see that if $\mathcal{M}^p$ has torsion cycles, we may get additional contributions. Concretely, we are interested in $p=q=5$ and $X^5=\mathbb{RP}^5$ whose cohomology groups with ordinary coefficients are well-known to be
\begin{equation}
	H^*(\mathbb{RP}^5;\mathbb{Z})=\{\mathbb{Z}, 0, \mathbb{Z}_2, 0, \mathbb{Z}_2, \mathbb{Z}\}\, .
\end{equation}
We also need the cohomology groups with local coefficients for fields which are twisted by the orientifold projection:
\begin{equation}
	H^*(\mathbb{RP}^5;\tilde{\mathbb{Z}})=\{0, \mathbb{Z}_2, 0, \mathbb{Z}_2, 0, \mathbb{Z}_2\}\, .
\end{equation}
We shall denote cocycles representing the generators for the cohomology groups with ordinary coefficients $u_n$ and those representing cohomology groups with local coefficients $t_n$. Their flat (if possible) uplifts to differential cocycles will be denoted $\breve{u}_n$ and $\breve{t}_n$ respectively.

We will now show how the above considerations modify our ansatz for dimensional reduction, starting with $\breve{H}_3$ and $\breve{F}_3$. These are acted upon by the orientifold projection so their expansion involves local coefficients. From the discussion in appendix \ref{sec:Kuenneth}, the most general form of the characteristic class is $I(\breve{F}_3)=[c_2\smile t_1 + p_3\smile h_0 + s_2\smile t_1]$. Here, $c_2\in Z^2(\mathcal{M}^5;\mathbb{Z})$. $p_3\in Z^3(\mathcal{M}^5;\mathbb{Z})$ is assumed to represent a 2-torsion cohomology class so $2p_3=\delta s_2$. Moreover, $[t_1]$ is torsion so $2t_1=\delta h_0$. From this, we may write down the flat differential cocycle 
$
	\breve{F}_3=((c_2+s_2)\smile t_1 + p_3\smile h_0, -\frac{1}{2}(c_2+s_2)\smile h_0,0)
$.
Observe now that $c_2+s_2$ is the integer uplift of a  $\mathbb{Z}_2$ cocycle, which we will call $c_{D1}$. Denote the integer uplift by $\tilde{c}_{D1}$. So every flat differential character can be described by a $\mathbb{Z}_2$ gauge field. Conversely, we could have started with an arbitrary $\mathbb{Z}_2$ cocycle $c_{D1}$ and defined
\begin{equation}
	\breve{F}_3=(\tilde{c}_{D1}\smile t_1 + \frac{1}{2}\delta\tilde{c}_{D1}\smile h_0, -\frac{1}{2}\tilde{c}_{D1}\smile h_0,0)
\end{equation}
which is a valid Hopkins-Singer cocycle. This shows that after dimensional reduction, we get a $\mathbb{Z}_2$ gauge field on $\mathcal{M}^5$. We similarly have
\begin{equation}
	\breve{H}_3=(\tilde{b}_{F1}\smile t_1 + \frac{1}{2}\delta\tilde{b}_{F1}\smile h_0, -\frac{1}{2}\tilde{b}_{F1}\smile h_0,0)\, .
\end{equation}
Of course, a similar analysis can be performed for all other supergravity fields, however none of them will lead to additional terms in the symmetry TFT so we shall not go through the exercise here.

\subsection{Symmetry TFT couplings}

Focus now on the supergravity coupling $2\pi i\int_{\mathcal{M}^5\times\mathbb{RP}^5}\breve{F}_5\star\breve{H}_3\star\breve{F}_3$. We use the expansion  of $\breve{F}_5$ given in \cite{GarciaEtxebarria:2022vzq}:
\begin{equation}
    \breve{F}_5=N\breve{1}+\breve{a}_1\star\breve{u}_4+\breve{a}_3\star\breve{u}_2+N\breve{u}_5
\end{equation}
for a stack of $N$ D3 branes on the $\mathbb{RP}^5$.

As mentioned  before, the discussion of the previous section should apply to this field as well, leading to additional terms. However, these terms do not end up affecting our results in this case. We get two contributions to the symmetry TFT: the first is due to the $\breve{a}_1\star\breve{u}_4$ term and leads to a $\pi i \int_{\mathcal{M}^5}a_1\smile b_{F1}\smile c_{D1}$ term in the symmetry TFT as calculated in \cite{Etheredge:2023ler}.
The second contribution comes from the $N\breve{u}_5$ term. Without loss of generality, we take the Hopkins-Singer cocycle $\breve{u}_5=(u_5,0,\text{vol}_{\mathbb{RP}^5})$. We compute:
\begin{equation}
	\breve{u}_5\star\breve{H}_3\star\breve{F}_3=(0,\frac{1}{4}u_5\smile\delta\tilde{b}_{F1}\smile h_0\smile\tilde{c}_{D1}\smile h_0,0)\, .
\end{equation}
Fibre-integration over $\mathbb{RP}^5$ then leads to a term $\pi i \int_{\mathcal{M}^5}\frac{N}{2}\delta \tilde{b}_{F1}\smile \tilde{c}_{D1}$ in the symmetry TFT. This matches our expectations from the anomaly theory of SO$(2N)_+$ gauge theory \cite{Hsin_2021}.

In addition to these anomaly terms, the symmetry TFT contains $BF$ terms. In \cite{Etheredge:2023ler}, these were deduced from the non-commutativity of branes in the bulk. To derive them directly from the supergravity action requires a manifestly self-dual formulation, such as the one proposed in \cite{Belov:2006jd,Belov:2006xj}, where the 10-dimensional action arises as an edge mode of an 11-dimensional bulk theory. This was applied to the calculation of symmetry TFTs from string theory in \cite{GarciaEtxebarria:2024fuk}. We shall not carry out a full analysis using these techniques here but the general story is that the IIB kinetic terms lead to $BF$ terms in the symmetry TFT. For example, we may decompose $\breve{F}_7$, the magnetic dual of $\breve{F}_3$, as $\breve{F}_7=\breve{c}_{D5}\star\breve{t}_5$. The kinetic term $\int_{\mathcal{M}^5\times\mathbb{RP}^5}\text{R}( \breve{F}_3)\wedge \text{R}( \breve{F}_7)$ then leads to the $BF$ term $\int_{\mathcal{M}^5}c_{D1}\smile \delta c_{D5}$ after dimensional reduction.

The total action for the symmetry TFT is then:
\begin{equation}
	S_{5D}=i\pi\int_{\mathcal{M}^5}a_1\smile \delta a_3+b_{F1}\smile\delta b_{NS5}+c_{D1}\smile\delta c_{D5}+a_1\smile b_{F1}\smile c_{D1}+\frac{N}{2}\delta b_{F1}\smile c_{D1}\, .
\end{equation}
 \subsection*{Stacking TFTs on symmetry generators}
 The na\"ive topological operators of the symmetry TFT are the  holonomies of the discrete gauge fields. However, in many cases these are not gauge invariant and must be stacked with a topological field theory on the defect worldvolume. It was pointed out in \cite{GarciaEtxebarria:2022vzq,Apruzzi:2022rei,Heckman:2022muc,Heckman:2022xgu,Apruzzi:2023uma,Bah:2023ymy} that this topological action can be computed from the WZ couplings on the brane worldvolume. We shall see that while the TFT on the fivebrane worldvolumes calculated in \cite{GarciaEtxebarria:2022vzq} are correct for $\mathfrak{so}(8k)$ theories, in the $\mathfrak{so}(8k+2)$ case, there are additional terms, which were first calculated from field theory considerations in \cite{Bhardwaj_2023i}.
 
 The couplings for D3 branes wrapping $\mathbb{RP}^1\subset\mathbb{RP}^5$ were calculated to be 
 \begin{equation}
 	D3(M^3)=\int \mathcal{D}\gamma_1\mathcal{D}\phi_1\exp(i \pi \int_{M^3}a_3+\gamma_1\smile\delta\phi_1+c_{D1}\smile\gamma_1+b_{F1}\smile c_{D1})
 \end{equation}
 and for D5 branes in the $N=4k$ case:
 \begin{equation}
 	D5(\Sigma^2)=\mathcal{D}\gamma_1\mathcal{D}\phi_0\exp(i \pi \int_{\Sigma^2}c_{D5}+\gamma_1\smile\delta\phi_0+\phi_0\smile b_{F1}+\gamma_1\smile a_1)\, .
 \end{equation}
 The action for the NS5 follows by S-duality. In general, the above discussion leads to additional terms for the D5 generator: a D5 wrapping $\Sigma_2\times\mathbb{RP}^4$, where $\mathbb{RP}^4$ represents a cocycle of $\mathbb{RP}^5$ with local coefficients, has a coupling $-\frac{1}{2}\int_{C^7}\Breve{H}_3\star\Breve{F}_5$ where $\partial C^7=\Sigma^2\times\mathbb{RP}^4$ \cite{Etheredge:2023ler}. We compute
 \begin{equation}
 	-\frac{1}{2}\int_{C^7}\Breve{H}_3\star\Breve{F}_5=-\frac{1}{2}\left(\int_{\Sigma^2\times\Tilde{pt}}\Breve{H}_3\right)\left(\int_{\mathbb{RP}^4}F_5\right)=-\frac{N}{2}\int_{\Sigma^2}b_{F1}\, .
 \end{equation}
 On the brane worldvolume, we have a $U(1)$ gauge field $\breve{F}_2$. From the above discussion, the most general ansatz is 
 \begin{equation}
 	\Breve{F}_2=(\tilde{\gamma}_1\smile t_1-\frac{1}{2}\delta\tilde{\gamma}_1\smile h_0,\frac{1}{2}(\tilde{\gamma}_1\smile h_0),0)\, .
 \end{equation}
 Here, $\gamma_1\in C^2(\Sigma_2;\mathbb{Z}_2)$ and $\tilde{\gamma}_1$ denotes an integer uplift. By a slight abuse of notation, $t_1$ is a representative of the generator of $H^1(\mathbb{RP}^4;\tilde{\mathbb{Z}})$. The worldvolume coupling $\breve{F}_2\star\breve{F}_5$ then contributes
 \begin{equation}
 	\int_{\Sigma_2}\gamma_1\smile a_1+\frac{N}{2}\delta\tilde{\gamma}_1\, .
 \end{equation}
 Thus, the gauge-invariant defect for general $N$ is
 \begin{equation}\label{D5action}
 		D5(\Sigma^2)=\mathcal{D}\gamma_1\mathcal{D}\phi_0\exp(i \pi \int_{\Sigma^2}c_{D5}+\gamma_1\smile\delta\phi_0+\phi_0\smile b_{F1}+\gamma_1\smile a_1+\frac{N}{2}(\delta\tilde{\gamma}_1-\tilde{b}_{F1}))
 \end{equation}
 which agrees with the results in \cite{Bhardwaj_2023} for odd $N$.
\section{Gapped boundaries and discrete theta angles}
The basic construction we are considering is the ``sandwich" \cite{Apruzzi:2021nmk,Gaiotto:2020iye,bhardwaj2023generalizedchargesiinoninvertible}. Here, the symmetry TFT is compactified on an interval with gapped boundary conditions on one boundary (the ``topological boundary") and gapless boundary conditions on the other (the ``physical boundary"). All global information, such as the global structure of the gauge group and discrete theta angles, is encoded in the topological boundary. In simple cases, topological boundary conditions in 3D TFT correspond to Lagrangian subgroups of the defect group \cite{Kapustin_2011} by allowing the line operators in such a subgroup to end on the boundary. More recently, gapped boundary conditions of 4D TFT have been studied systematically \cite{bhardwaj2024gappedphases21dnoninvertible,Bullimore:2024khm}. 

We are not aware of precise results of this nature for the 5D case, but following the approach of \cite{Kaidi:2023maf} we will assume that a gapped boundary condition is specified by a set of operators such that all linking numbers and higher linking numbers between them vanish. In general, this is only a necessary but not a sufficient condition to define a valid boundary condition. However, in this case we will see that each boundary condition we find can be matched to a global form, so it appears to be sufficient as well. 

\subsection{$\mathfrak{so}(8k)$ theories}
This case is the simplest one because the $\frac{N}{2}$ term does not contribute to the action.\footnote{The discussion in this section applies to all $k>1$, however for $k=1$, the 0-form symmetry of Spin$(8)$ gauge theory is $D_3$, the symmetry group of the Dynkin diagram \cite{Kaidi_2022}. We do not attempt to identify the brane corresponding to the order 3 generator of this symmetry. Indeed, since we are now at small $N$, we cannot trust the holographic computation. However it is interesting to consider how this tentative operator acts on the lines of the theory.
This generator corresponds to a rotation of the Dynkin diagram by $120$ degrees. As such, it sends a Wilson line in the s-spinor representation, given by a D5 fat string to a Wilson line in the vector representation, given by an F1 string. This suggests that this symmetry generator treats F1 and D5 charges on the same footing. From the point of view of cohomology, this does not seem to make sense. If however we replace cohomology by some kind of S-duality covariant K-theory, we expect that both brane charges should live in the same group. The role of K-theory in generalised symmetries was investigated in \cite{Zhang:2024oas}.}  The boundary conditions for which the 0-form symmetry is not gauged (i.e. Dirichlet for $a_1$) have already been identified in \cite{Etheredge:2023ler,Bergman_2022} and can be seen to agree with \cite{Aharony:2013hda}.

To match the boundary conditions to global forms, we need to choose a duality frame. We shall use the same conventions as \cite{Etheredge:2023ler} and fix the generators 
\begin{align}
	\begin{aligned}\label{dualitytransform}
		b_{F1}&\xrightarrow{\text{S}}-c_{D1},\quad
		&c_{D1}&\xrightarrow{\text{S}}b_{F1},\quad
		&b_{NS5}&\xrightarrow{\text{S}}-c_{D5},\quad
		&c_{D5}&\xrightarrow{\text{S}}b_{NS5},\\
		b_{F1}&\xrightarrow{\text{T}}b_{F1},\quad
		&c_{D1}&\xrightarrow{\text{T}}c_{D1}-b_{F1},\quad
		&b_{NS5}&\xrightarrow{\text{T}}b_{NS5}+c_{D5},\quad
		&c_{D5}&\xrightarrow{\text{T}}c_{D5}\, .
	\end{aligned}
\end{align}
S-duality leaves $F_5$ and hence $a_1$ and $a_3$ invariant.
Moreover, we fix the global form Spin$(8k)$ to correspond to the Dirichlet boundary conditions 
\begin{align}
	\begin{aligned}			b_{F1}\Bigr|_{\substack{\partial\mathcal{M}^5}}=B_2^{(v)}\qquad&&c_{D5}\Bigr|_{\substack{\partial\mathcal{M}^5}}=B_2^{(c)}\qquad&&a_1\Bigr|_{\substack{\partial\mathcal{M}^5}}=A_1\, .
	\end{aligned}
\end{align}
In other words, F1 strings on a point in $\mathbb{RP}^5$, D5 branes wrapping a twisted cocycle represented by $\mathbb{RP}^4\subset\mathbb{RP}^5$, and D3 branes wrapping $\mathbb{RP}^3\subset\mathbb{RP}^5$ can end on the boundary. In field theory, the background fields correspond to Stiefel-Whitney classes. Concretely, a PO$(8k)$ bundle has Stiefel-Whitney classes $w_1$, $w_2^{(s)}$, and $w_2^{(c)}$. See \cite{Hsin_2021} for a physics-oriented review and \cite{kirby1990pin} for a rigorous mathematical treatment. We define $w_2^{(v)}=w_2^{(s)}+w_2^{(c)}$. The Stiefel-Whitney classes satisfy the relation $\delta w_2^{c}=w_1\smile w_2^{(v)}$. This gives us the 2-group structure of the Spin$(8k)$ theory \cite{Bhardwaj_2023}.

\sloppy We can now turn on backgrounds for the symmetries of the Spin$(8k)$ by fixing the values of these Stiefel-Whitney classes. When computing the partition function $Z_{\text{Spin}(8k)}(\tau,B_2^{(s)},B_2^{(c)},A_1)$, we only sum over PO$(8k)$-bundles with Stiefel-Whitney classes given by $w_1=B_2^{(s)}$, $w_2^{(s)}=B_2^{(s)}$, and $w_2^{(c)}=B_2^{(c)}$ and then perform the path integral over connections on these bundles. In particular if we turn all backgrounds off, we sum exactly over those PO$(8k)$ bundles which can be lifted to Spin$(8k)$ bundles.

As a warm-up exercise, we consider the case of O$(8k)_{++}$. One can obtain this theory by gauging the 0-form symmetry of SO$(8k)_+$. As suggested by the notation above, the O$(8k)$ gauge theory has an additional discrete theta angle, which the SO$(8k)$ theory does not possess \cite{Hsin_2021}. Gauging the 0-form symmetry results in a dual 2-form symmetry. The only bulk field which could be a background for this is $a_3$, so we know that giving this field a Dirichlet boundary condition is equivalent to gauging the 0-form symmetry. Since the 1-form symmetry is not being gauged, we expect to have the same boundary conditions for the 2-form gauge fields as in the O$(8k)_+$ case. In summary:
\begin{align}
 	\begin{aligned}
 		b_{F1}\Bigr|_{\substack{\partial\mathcal{M}^5}}=B_2^e \qquad&&c_{D1}\Bigr|_{\substack{\partial\mathcal{M}^5}}=B_2^m\qquad&&a_3\Bigr|_{\substack{\partial\mathcal{M}^5}}=A_3\, .
 	\end{aligned}
\end{align}
To see that this matches our expectation, let us define the physical boundary condition following \cite{Kaidi:2022cpf}:
\begin{equation}
	|\chi \rangle=\sum_{b_{F1},c_{D5},a_1}Z_{Spin(8k)}(\tau,b_{F1}+c_{D5},c_{D5},a_1)|b_{F1},c_{D5},a_1\rangle\, .
\end{equation}
Here, we chose $Z_{Spin(8k)}$ as our reference point since the background fields are simply the Stiefel-Whitney classes in this case. The gapped boundary condition described above is then implemented by a state
\begin{equation}
	\sum_{b_{F1},c_{D5},a_1}\delta(b_{F1}-B_2^e)(-1)^{\int_{\mathcal{M}^4}c_{D5}\smile B_2^m}(-1)^{\int_{\mathcal{M}^4}a_1\smile A_3}|b_{F1},c_{D5},a_1\rangle\, .
\end{equation}
The sandwich construction instructs us to compute the inner product of these boundary conditions, yielding
\begin{align}
	\begin{split}
			&\sum_{b_{F1},c_{D5},a_1}\langle b_{F1},c_{D5},a_1|\delta(b_{F1}-B_2^e)(-1)^{\int_{\mathcal{M}^4}c_{D5}\smile B_2^m}(-1)^{\int_{\mathcal{M}^4}a_1\smile A_3}|\chi\rangle\\
			=&\sum_{c_{D5}}Z_{Spin(8k)}(\tau,B_2^e+c_{D5},c_{D5},a_1)(-1)^{\int_{\mathcal{M}^4}c_{D5}\smile B_2^m}\\
			=&\,Z_{O(8k)_+}(\tau,B_2^e,B_2^m,A_3)\, .
	\end{split}
\end{align}
We see that when $B_2^e=0$, we are summing PO bundles with $w_2^{(v)}=0$, which is precisely the condition for lifting to an O-bundle. As expected from the SO-case, the symmetry boundary condition is invariant under S-duality. From the equations of motion of the symmetry TFT, we find that $\delta a_3=b_{F1}\smile c_{D1}$, also consistent with the modified Bianchi identity for $F_5$. In terms of the backgrounds, this means we have a 3-group whose background fields obey $\delta A_3=B_2^e\smile B_2^m$ in agreement with \cite{Hsin_2021}.

Let us now add a discrete theta angle to the discussion. By similar reasoning as before, we expect the boundary condition for O$(8k)_{-+}$ to be 
\begin{align}
	\begin{aligned}
		b_{F1}\Bigr|_{\substack{\partial\mathcal{M}^5}}=B_2^e \qquad&&c_{D1}-c_{D5}\Bigr|_{\substack{\partial\mathcal{M}^5}}=B_2^m\qquad&&a_3\Bigr|_{\substack{\partial\mathcal{M}^5}}=A_3
	\end{aligned}
\end{align}
implemented by 
\begin{equation}
		\sum_{b_{F1},c_{D5},a_1}\delta(b_{F1}-B_2^e)(-1)^{\int_{\mathcal{M}^4}c_{D5}\smile B_2^m+\frac{1}{2}\mathcal{P}(c_{D5})}(-1)^{\int_{\mathcal{M}^4}a_1\smile A_3}|b_{F1},c_{D5},a_1\rangle\, .
\end{equation}

A similar calculation shows that this is indeed O$(8k)_{-+}$ gauge theory. The extension of $B_2^m$ to the bulk is now $c_{D1}-c_{D5}$ rather than $c_{D1}$. So we rewrite the anomaly term in the symmetry TFT as
\begin{equation}
	\int_{\mathcal{M}^5}a_1\smile b_{F1}\smile c_{D1}=\int_{\mathcal{M}^5}a_1\smile b_{F1}\smile (c_{D1}-c_{D5})+a_1\smile b_{F1}\smile c_{D5}\, .
\end{equation}
This matches the anomaly theory found in \cite{Hsin_2021}:
\begin{equation}
	\int_{\mathcal{M}^5}w_1\smile B_2^e\smile B_2^m+w_1\smile B_2^e\smile w_2^{(c)}\, .
\end{equation}
We also know that $\delta a_3 = b_{F1}\smile c_{D1}$ so on the boundary we have $\delta A_3 = B_2^e\smile B_2^m + B_2^e\smile c_{D5}$. We see that we cannot consistently turn on $B_2^e$. It turns out that the electric 1-form symmetry becomes non-invertible in this case as explained in \cite{Hsin_2021}.
 The above boundary condition is invariant under a $T$ transformation but an $S$ transformation takes it to 
\begin{align}
	\begin{aligned}
		c_{D1}\Bigr|_{\substack{\partial\mathcal{M}^5}}=B_2^e \qquad&&b_{F1}-b_{NS5}\Bigr|_{\substack{\partial\mathcal{M}^5}}=B_2^m\qquad&&a_3\Bigr|_{\substack{\partial\mathcal{M}^5}}=A_3\, .
	\end{aligned}
\end{align}
One may again compute the partition function, revealing that we are now summing over PO-bundles with a discrete theta term $\frac{1}{2}\mathcal{P}(w_2^{(v)})$. We will call this theory PO$(8k)_{\begin{smallmatrix}-&-&+\\-&-&\end{smallmatrix}}$ anticipating the appearance of a new discrete theta angle. One can cross-check this result using what is known about S-duality in the connected case. We have

\begin{flalign}
	&\begin{aligned}
		&Z_{\text{O}(8k)_{-+}}(\tau,B_2^e,B_2^m,A_3)&\\
		&\qquad=\sum_{w_1,w_2^{(c)}}Z_{\text{Spin}(8k)}(\tau,B_2^e+w_2^{(c)},w_2^{(c)},w_1)(-1)^{\int_{\mathcal{M}^4}B_2^m\smile w_2^{(c)}+\frac{1}{2}\mathcal{P}(w_2^{(c)})}(-1)^{\int_{\mathcal{M}^4}A_3\smile w_1}	\\
		&\qquad\stackrel{\text{S}}{=}\sum_{a_1,b_2}Z_{\text{PSO}(8k)_{\begin{smallmatrix}+&+\\+&+\end{smallmatrix}}}(-\frac{1}{\tau},B_2^e+b_2,b_2,a_1)(-1)^{\int_{\mathcal{M}^4}B_2^m\smile b_2+\frac{1}{2}\mathcal{P}(b_2)}(-1)^{\int_{\mathcal{M}^4}A_3\smile a_1}	\\
		&\qquad=Z_{\text{PO}_{\begin{smallmatrix}-&-&+\\-&-&\end{smallmatrix}}}(-\frac{1}{\tau},B_2^m,B_2^e,A_3)
	\end{aligned}&&
\end{flalign}

where the final step follows by expanding $Z_{\text{PSO}(8k)_{\begin{smallmatrix}+&+\\+&+\end{smallmatrix}}}$ in terms of $Z_{\text{Spin}(8k)}$ and performing the sum. The action of the $T$ generator takes this boundary condition to
\begin{align}
	\begin{aligned}
		b_{F1}+c_{D1}\Bigr|_{\substack{\partial\mathcal{M}^5}}=B_2^{(1)} \qquad&&b_{F1}-b_{NS5}+c_{D5}\Bigr|_{\substack{\partial\mathcal{M}^5}}=B_2^{(2)}\qquad&&a_3\Bigr|_{\substack{\partial\mathcal{M}^5}}=A_3
	\end{aligned}
\end{align}
which corresponds to the PO$(8k)_{\begin{smallmatrix}+&-&+\\-&+&\end{smallmatrix}}$ theory.

There is one more duality orbit to consider, which is the one containing the Spin$(8k)$ theory. Gauging the 0-form symmetry will take us to the Pin$^+(8k)$ theory. Since we know the boundary condition for the Spin$^(8k)$ theory, we claim that the boundary condition for Pin$^+$ is

\begin{align}
	\begin{aligned}
		b_{F1}\Bigr|_{\substack{\partial\mathcal{M}^5}}=B_2^{(1)} \qquad&&c_{D5}\Bigr|_{\substack{\partial\mathcal{M}^5}}=B_2^{(2)}\qquad&&a_3\Bigr|_{\substack{\partial\mathcal{M}^5}}=A_3\, .
	\end{aligned}
\end{align}
One can again check that the sandwich construction results in $Z_{\text{Spin}(8k)}(\tau, B_2^{(1)}+B_2^{(2)},B_2^{(2)},A_3)$. The anomaly term in the symmetry TFT tells us that there is a gauge-global anomaly between $B_2^{(1)}$, $a_1$, and $c_{D1}$. Again, this suggests that the former symmetry becomes non-invertible. We can complete the duality orbit as before. An $S$ transformation takes us to

\begin{align}
	\begin{aligned}
		c_{D1}\Bigr|_{\substack{\partial\mathcal{M}^5}}=B_2^{(1)} \qquad&&b_{NS5}\Bigr|_{\substack{\partial\mathcal{M}^5}}=B_2^{(2)}\qquad&&a_3\Bigr|_{\substack{\partial\mathcal{M}^5}}=A_3
	\end{aligned}
\end{align}
corresponding to PO$(8k)_{\begin{smallmatrix}+&+&+\\+&+&\end{smallmatrix}}$. From here, $T$ results in 
\begin{align}
	\begin{aligned}
		b_{F1}+c_{D1}\Bigr|_{\substack{\partial\mathcal{M}^5}}=B_2^{(1)} \qquad&&c_{D5}+b_{NS5}\Bigr|_{\substack{\partial\mathcal{M}^5}}=B_2^{(2)}\qquad&&a_3\Bigr|_{\substack{\partial\mathcal{M}^5}}=A_3
	\end{aligned}
\end{align}
i.e. the PO$(8k)_{\begin{smallmatrix}-&+&+\\+&-&\end{smallmatrix}}$ theory. So far, this has been a rather straightforward exercise; the duality orbits we found mirror those determined in \cite{Aharony:2013hda}. It is known that there are duality orbits which are unique to $\mathfrak{so}(8k)$ or $\mathfrak{so}(8k+4)$ theories. However it turns out that for those theories, one cannot gauge the 0-form symmetry so they do not enter our discussion. The structures resulting from these considerations are shown in figure \ref{figure:so8k1}.
\subsection*{Obstruction to other gapped boundary conditions}
We know that the symmetry TFT contains a coupling $b_{F1}\smile c_{D1}\smile a_1$. A calculation similar to \cite{Kaidi:2023maf} shows that this results in a D5 fat string, an NS5 fat string, and a D3 wrapping $\mathbb{RP}^1\subset\mathbb{RP}^5$ link nontrivially.\footnote{Such a link is called a three component link of type 0. Let the worldvolumes of the D5 fat string, the NS5 fat string, and the wrapped D5 be $M_1^{(2)}$, $M_2^{(2)}$, and $M_3^{(3)}$. Assuming that they wrap trivial cycles, let their Seifert surfaces be $N_1^{(3)}$, $N_2^{(3)}$, and $N_3^{(4)}$. One defines $\text{Link}(M_1^{(2)},M_2^{(2)},M_3^{(3)}):=\int\text{PD}(N_1^{(3)})\smile \text{PD}(N_2^{(3)})\smile \text{PD}(N_3^{(4)})$. See \cite{doi:10.1142/S0218216598000206} for mathematical details and \cite{Putrov_2017,Wan:2019oyr,Zhang_2021,Zhang_2022,Del_Zotto_2024} for a selection of applications in physics.} This means that there is no gapped boundary condition where those three objects can all end on the boundary. This matches our expectation since such a boundary condition would correspond to gauging the non-invertible 0-form symmetry of Sc$(8k)_+$. However, one obtains Sc$(8k)_+$ by gauging the $\mathbb{Z}_2^{(e)}\times \mathbb{Z}_2^{(m)}$ 1-form symmetry of SO$(8k)_+$ so gauging the 0-form of Sc$(8k)_+$ would be equivalent to gauging the entire symmetry of SO$(8k)_+$. This is not possible because of the mixed 't Hooft anomaly of the SO$(8k)_+$ theory. Similar arguments tell us that we cannot gauge the 0-form symmetry in the remaining duality orbits either.
\subsubsection{Symmetry fractionalisation}

The case of Pin$^-(8k)$ is more subtle. A PO$(8k)$ bundle can be lifted to a Pin$^-(8k)$ bundle iff its Stiefel-Whitney classes satisfy the relation \cite{kirby1990pin}
\begin{equation}
	w_2^{(v)}+w_1^2=0\, .
\end{equation}
We arrive at this theory by gauging the 2-subgroup $\mathbb{Z}_2^{(v)}\times\mathbb{Z}_2^{(0)}$ with nontrivial symmetry fractionalisation as sketched in \cite{Hsin_2021}. This makes sense because this 2-subgroup has trivial Postnikov class. Intuitively, we are summing over networks of 0-form symmetry defects with 1-form symmetry defects generating the diagonal $\mathbb{Z}_2^{(v)}$ sitting at each junction. This is the case because the cup product can roughly be understood as the Poincar\'{e} dual of the intersection product. Putting all of this together, the partition function of the Pin$^-(8k)$ theory is given by

\begin{multline}\label{pinminus}
	Z_{\text{Pin}^-(8k)_+}(\tau,B_2^e,B_2^m,A_3)=\sum_{\substack{w_1\\w_2^{(c)}}}Z_{\text{Spin}(8k)}(\tau,w_2^{(c)}+(w_1)^2+B_2^{e},w_2^{(c)},w_1)\\(-1)^{\int_{\partial\mathcal{M}^5}w_2^{(c)}\smile B_2^m}(-1)^{\int_{\partial\mathcal{M}^5}w_1\smile A_3}\, .
\end{multline}
As expected, for $B_2^e=0$, we are summing exactly over those PO$(8k)$ bundles which can be lifted to Pin$^-(8k)$ bundles. We can include a discrete theta angle given by $\pi i\int_{\partial\mathcal{M}^5}(w_1)^2\smile w_2^{(c)}$ in the partition function. We denote the theory ``without" such a discrete theta angle by Pin$^-(8k)_+$ and the one ``with" discrete theta angle by Pin$^-(8k)_+$.\footnote{As we shall see, these theories are actually related by S-duality so talking about global forms with/without the discrete theta angle only makes sense after choosing a duality frame.}

Consider now the boundary condition given by
\begin{align}
	\begin{aligned}
	b_{F1}-(a_1)^2\Bigr|_{\substack{\partial\mathcal{M}^5}}=B_2^e \qquad&&c_{D1}\Bigr|_{\substack{\partial\mathcal{M}^5}}=B_2^m\qquad&&a_3\Bigr|_{\substack{\partial\mathcal{M}^5}}=A_3
	\end{aligned}
\end{align}
 This is a natural guess since the first boundary condition is reminiscent of the condition $w_2^{(v)}+(w_1)^2=B_2^e$ and in the Spin$(8k)$ case, we identified $b_{F1}$ with $w_2^{(v)}$ and $a_1$ with $w_1$ on the boundary. All (higher) linking invariants of the associated operators vanish so we expect this to define a Lagrangian algebra and therefore a gapped boundary condition. Indeed, we compute
 
 \begin{align}
 	\begin{split}
 		&\sum_{b_{F1}, c_{D5}, a_1}	\langle b_{F1},c_{D5},a_1 |\delta(b_{F1}-(a_1)^2-B_2^e)(-1)^{\int_{\partial\mathcal{M}^5}c_{D5}\smile B_2^m}\delta(a_3-A_3)| \chi \rangle\\
 		=&\sum_{\substack{a_1\\c_{D5}}}Z_{\text{Spin}(8k)}(\tau,c_{D5}+(a_1)^2+B_2^{e},c_{D5},a_1)(-1)^{\int_{\partial\mathcal{M}^5}c_{D5}\smile B_2^m}(-1)^{\int_{\partial\mathcal{M}^5}a_1\smile A_3}\\
 		=&\,	Z_{\text{Pin}^-(8k)_+}(\tau,B_2^e,B_2^m,A_3)\, .
 	\end{split} 
 \end{align}
 We are now interested in the duality orbit in which this boundary condition lies. The $S$ generator sends it to 
 
 \begin{align}
 	\begin{aligned}
 		b_{F1}\Bigr|_{\substack{\partial\mathcal{M}^5}}=B_2^e \qquad&&c_{D1}-(a_1)^2\Bigr|_{\substack{\partial\mathcal{M}^5}}=B_2^m\qquad&&a_3\Bigr|_{\substack{\partial\mathcal{M}^5}}=A_3
 	\end{aligned}\, .
 \end{align}
 The state corresponding to this boundary condition is 
 \begin{equation}
 	\sum_{b_{F1}, c_{D5}, a_1}\delta(b_{F1}-B_2^e)(-1)^{\int_{\partial\mathcal{M}^5}c_{D5}\smile((a_1)^2+ B_2^m)}(-1)^{\int_{\partial\mathcal{M}^5}a_1\smile A_3}|b_{F1},c_{D5},a_1\rangle\, .
 \end{equation}
 As explained in the appendix of \cite{Lawrie:2023tdz}, this may be written in terms of states of well-defined $c_{D1}$ flux on the boundary as
 \begin{equation}
 	|b_{F1},c_{D5},a_1\rangle=\sum_{c_{D1}}(-1)^{\int_{\partial\mathcal{M}^5}c_{D5}\smile c_{D1}}|b_{F1},c_{D1},a_1\rangle\, .
 \end{equation}
 Substituting this into the previous expression, we see that the sum over $c_{D5}$ forces $c_{D1}+(a_1)^2+B_2^m=0$ as desired.

 We may again calculate the resulting partition function:
 
 \begin{align}
 	\begin{split}
 		&\sum_{b_{F1}, c_{D5}, a_1}	\langle b_{F1},c_{D5},a_1 |\delta(b_{F1}-B_2^e)(-1)^{\int_{\partial\mathcal{M}^5}c_{D5}\smile((a_1)^2+ B_2^m)}(-1)^{\int_{\partial\mathcal{M}^5}a_1\smile A_3}| \chi \rangle\\
 		=&\sum_{\substack{a_1\\c_{D5}}}Z_{\text{Spin}(8k)}(-\frac{1}{\tau},c_{D5}+B_2^{e},c_{D5},a_1)(-1)^{\int_{\partial\mathcal{M}^5}c_{D5}\smile (B_2^m+(a_1)^2)}(-1)^{\int_{\partial\mathcal{M}^5}a_1\smile A_3}\\
 		=&	\,Z_{\text{O}(8k)_{+-}}(-\frac{1}{\tau},B_2^e,B_2^m,A_3)\, .
 	\end{split}
 \end{align}
 Here, O$(8k)_{+-}$ denotes O$(8k)$ gauge theory with discrete theta angle $\pi\int_{\mathcal{M}^4}w_2^{(c)}\smile (w_1)^2$ turned on. This global form and some of its properties were discussed in \cite{Hsin_2021} and we shall make contact with the discussion therein shortly. Before we do so, we shall show that this is indeed how we expect the $S$ generator to act. We know that Spin$(8k)$ SYM is S-dual to PSO$(8k)_{\begin{smallmatrix}+&+\\+&+\end{smallmatrix}}$ \cite{Aharony:2013hda}. Since we can obtain Pin$^-(8k)_+$ by gauging Spin$(8k)$ with an appropriate symmetry fractionalisation, we may leverage this to find its S-dual.\footnote{The author thanks Jamie Pearson for discussions related to this issue.} Starting with our partition function for Pin$^-(8k)_+$, we get

  \begin{flalign}
 	&\begin{aligned}
 		&Z_{\text{Pin}^-(8k)_+}(\tau,B_2^e,B_2^m,A_3)\\&\qquad=\sum_{\substack{w_1\\w_2^{(c)}}}Z_{\text{Spin}(8k)}(\tau,w_2^{(c)}+(w_1)^2+B_2^{e},w_2^{(c)},w_1)(-1)^{\int_{\partial\mathcal{M}^5}w_2^{(c)}\smile B_2^m}(-1)^{\int_{\partial\mathcal{M}^5}w_1\smile A_3}\\
 		&\qquad\overset{\text{S}}{=}\sum_{\substack{a_1\\b_2}}Z_{\text{PSO}(8k)_{\begin{smallmatrix}+&+\\+&+\end{smallmatrix}}}(-\frac{1}{\tau},b_2+(a_1)^2+B_2^{e},b_2,a_1)(-1)^{\int_{\partial\mathcal{M}^5}b_2\smile B_2^m}(-1)^{\int_{\partial\mathcal{M}^5}a_1\smile A_3}\\
 		&\qquad=\sum_{\substack{w_1\\b_2}}\sum_{\substack{w_2^{(s)},w_2^{(c)}}}Z_{\text{Spin}(8k)}(-\frac{1}{\tau},w_2^{(s)},w_2^{(c)},w_1)(-1)^{\int_{\partial\mathcal{M}^5}w_2^{(s)}\smile (b_2+(w_1)^2+B_2^{e})}\\&\qquad\qquad(-1)^{\int_{\partial\mathcal{M}^5}b_2\smile w_2^{(c)}}(-1)^{\int_{\partial\mathcal{M}^5}b_2\smile B_2^m}(-1)^{\int_{\partial\mathcal{M}^5}w_1\smile A_3}\\
 		&\qquad=\sum_{\substack{w_1\\w_2^{(s)}}}Z_{\text{Spin}(8k)}(-\frac{1}{\tau},w_2^{(s)},w_2^{(s)}+B_2^m,w_1)(-1)^{\int_{\partial\mathcal{M}^5}w_2^{(s)}\smile ((w_1)^2+B_2^{e})}(-1)^{\int_{\partial\mathcal{M}^5}w_1\smile A_3}\\
 		&\qquad=\,Z_{\text{O}(8k)_{+-}}(-\frac{1}{\tau},B_2^m,B_2^e)\, .
 	\end{aligned}&&
 \end{flalign}
 Note that this exchanges the electric and magnetic background symmetries, as is expected for S-duality.
 We shall now make contact with the discussion in \cite{Hsin_2021}. There, it is shown that the O$(8k)_{+-}$-theory possesses a gauge-global mixed anomaly given by $\pi\int_{\mathcal{M}^5}B_2^e\smile(w_1)^3$, which turns the electric 1-form symmetry into a non-invertible symmetry. We can obtain this anomaly from the symmetry TFT using the fact that $B_2^e$ is the restriction of $c_{F1}-(a_1)^2$ to the boundary. Therefore, we are instructed to rewrite the action in terms of this combination of fields. This should then match the gauged version of the anomaly theory found in \cite{Hsin_2021}. Indeed,
 \begin{align}\label{symftf2}
 	\begin{split}
 			S_{5D} = i\pi\int_{\mathcal{M}^5}&a_1\smile \delta a_3+b_{F1}\smile\delta b_{NS5}+c_{D1}\smile\delta c_{D5}+a_1\smile b_{F1}\smile c_{D1}\\
 			=i\pi\int_{\mathcal{M}^5}&a_1\smile \delta a_3+b_{F1}\smile\delta b_{NS5}+(c_{D1}-(a_1)^2)\smile\delta c_{D5}+(a_1)^2\smile\delta c_{D5}\\
 			&a_1\smile b_{F1}\smile (c_{D1}-(a_1)^2)+a_1\smile b_{F1}\smile (a_1)^2\, .
 	\end{split}
 \end{align}
We now have
\begin{align}
	\begin{split}
		i\pi\int_{\mathcal{M}^5}(a_1)^2\smile\delta c_{D5}&=i\pi\int_{\mathcal{M}^5}\delta((a_1)^2\smile c_{D5})-\delta(a_1)^2\smile c_{D5}\\
		&=i\pi\int_{\partial\mathcal{M}^5}(a_1)^2\smile c_{D5}-i\pi\int_{\mathcal{M}^5}(\delta a_1\smile a_1-a_1\smile \delta a_1)\smile c_{D5}\\
		&=i\pi\int_{\partial\mathcal{M}^5}(a_1)^2\smile c_{D5}+i\pi\int_{\mathcal{M}^5}(\delta(\delta a_1\smile_1 a_1)-\delta a_1\smile_1 \delta a_1)\smile c_{D5}\, .
	\end{split}
\end{align}
The boundary term exactly matches the discrete theta angle discussed in \cite{Hsin_2021} while the role played by the bulk terms involving the higher cup product is not entirely clear to us. The first three terms in \ref{symftf2} are $BF$ terms for the finite background and gauge fields on the boundary. The next term gives us the discrete theta angle on the boundary as we have just seen and the final two terms exactly match the gauge global anomalies $i\pi\int_{\mathcal{M}^5}B_2^e\smile B_2^m\smile w_1$ and $i\pi\int_{\mathcal{M}^5}B_2^e\smile (w_1)^3$.

Of course, acting on O$(8k)_{+-}$ with the $S$ transformation will take us back to Pin$^-(8k)_+$ and $T$ leaves it invariant. The action of $T$ on Pin$^-(8k)_+$ results in:

\begin{align}
	\begin{aligned}
		b_{F1}-(a_1)^2\Bigr|_{\substack{\partial\mathcal{M}^5}}=B_2^e \qquad&&c_{D1}-(a_1)^2\Bigr|_{\substack{\partial\mathcal{M}^5}}=B_2^m\qquad&&a_3\Bigr|_{\substack{\partial\mathcal{M}^5}}=A_3
	\end{aligned}
\end{align}
with partition function

\begin{align}
	\begin{split}
		&\sum_{b_{F1}, c_{D5}, a_1}	\langle b_{F1},c_{D5},a_1 |\delta(b_{F1}-(a_1)^2-B_2^e)(-1)^{\int_{\partial\mathcal{M}^5}c_{D5}\smile((a_1)^2+ B_2^m)}\delta(a_3-A_3)| \chi \rangle\\
		=&\sum_{\substack{a_1\\c_{D5}}}Z_{\text{Spin}(8k)}(-\frac{1}{\tau},c_{D5}+(a_1)^2+B_2^{e},c_{D5},a_1)(-1)^{\int_{\partial\mathcal{M}^5}c_{D5}\smile (B_2^m+(a_1)^2)}(-1)^{\int_{\partial\mathcal{M}^5}a_1\smile A_3}\\
		=&\,	Z_{\text{Pin}(8k)_{-}}(-\frac{1}{\tau},B_2^e,B_2^m,A_3)\, .
	\end{split}
\end{align}
We can again check this by leveraging the action of $T$ on the Spin$(8k)$ gauge theory. As explained in \cite{Kaidi_2022}, $T$ sends this theory to itself, stacked with an SPT phase given by $(-1)^{\int_{\mathcal{M}^5}B_2^{(s)}\smile B_2^{(c)}}$. Using this and the definition \ref{pinminus} reproduces this result. This boundary condition is invariant under $S$ transformation so this completes the duality orbit.

This provides evidence that boundary conditions mixing bulk fields of various degrees are physically relevant. There are more such boundary conditions and we shall see how they correspond to various global forms of the gauge group. But first, a comment is in order. In \cite{Etheredge:2023ler} it is shown how D-brane charges as measured in ordinary cohomology correspond to the charges of line operators identified in \cite{Aharony:2013hda}. In this case, we cannot make a similar statement because it is not clear which charge is associated to the $(a_1)^2$ term. It has been shown \cite{Moore:1999gb} that the K-theory classification of RR fluxes relates the topology of fluxes of different degrees to each other. Moreover,  it is not clear to us precisely which brane configuration can end on the boundary in this case since there is no coupling on e.g. the D1 worldvolume which would give rise to an $(a_1)^2$-term. We expect that a more careful treatment of fluxes in terms of K-theory might shed some light on both of these issues.

In light of the previous results, we are led to expect the boundary condition

\begin{align}
	\begin{aligned}
		b_{F1}\Bigr|_{\substack{\partial\mathcal{M}^5}}=B_2^e \qquad&&c_{D1}-c_{D5}-(a_1)^2\Bigr|_{\substack{\partial\mathcal{M}^5}}=B_2^m\qquad&&a_3\Bigr|_{\substack{\partial\mathcal{M}^5}}=A_3
	\end{aligned}
\end{align}
to correspond to O$(8k)_{--}$ gauge theory. Indeed, the partition function is 

 \begin{align}
	\begin{split}
		&\sum_{b_{F1}, c_{D5}, a_1}	\langle b_{F1},c_{D5},a_1 |\delta(b_{F1}-B_2^e)(-1)^{\int_{\partial\mathcal{M}^5}\frac{1}{2}\mathcal{P}(c_{D5})+c_{D5}\smile((a_1)^2+ B_2^m)}\delta(a_3-A_3)| \chi \rangle\\
		=&\sum_{\substack{a_1\\c_{D5}}}Z_{\text{Spin}(8k)}(-\frac{1}{\tau},c_{D5}+B_2^{e},c_{D5},a_1)(-1)^{\int_{\partial\mathcal{M}^5}\frac{1}{2}\mathcal{P}(c_{D5})+c_{D5}\smile (B_2^m+(a_1)^2)}(-1)^{\int_{\partial\mathcal{M}^5}a_1\smile A_3}\\
		=&\,	Z_{\text{O}(8k)_{--}}(-\frac{1}{\tau},B_2^e,B_2^m,A_3)\, .
	\end{split}
\end{align}
The anomaly theory for O$(8k)_{--}$ gauge theory consists of terms we have already identified in the O$(8k)_{-+}$ and O$(8k)_{+-}$ cases, so we shall not go through the exercise again. But as before, the manipulations of the symmetry TFT do reproduce the anomaly terms expected from \cite{Hsin_2021}. This boundary condition is invariant under the $T$ transformation. The $S$ transformation sends it to
\begin{align}
	\begin{aligned}
		c_{D1}\Bigr|_{\substack{\partial\mathcal{M}^5}}=B_2^{(2)} \qquad&&b_{F1}+b_{NS5}-(a_1)^2\Bigr|_{\substack{\partial\mathcal{M}^5}}=B_2^{(1)}\qquad&&a_3\Bigr|_{\substack{\partial\mathcal{M}^5}}=A_3\, .
	\end{aligned}
\end{align}
As before, we compute
 \begin{align}
	\begin{split}
		&\sum_{b_{F1}, c_{D5}, a_1}	\langle b_{F1},c_{D5},a_1 |(-1)^{\int_{\partial\mathcal{M}^5}\frac{1}{2}\mathcal{P}(b_{F1})+b_{F1}\smile(B_2^{(1)}+(a_1)^2)}(-1)^{\int_{\partial\mathcal{M}^5}c_{D5}\smile B_2^{(2)}}\delta(a_3-A_3)| \chi \rangle\\
		=&\sum_{\substack{a_1\\c_{D5}}}Z_{\text{Spin}(8k)}(-\frac{1}{\tau},b_{F1},c_{D5},a_1)(-1)^{\int_{\partial\mathcal{M}^5}\frac{1}{2}\mathcal{P}(b_{F1})+b_{F1}\smile(B_2^{(1)}+(a_1)^2)}(-1)^{\int_{\partial\mathcal{M}^5}c_{D5}\smile B_2^{(2)}}\\&(-1)^{\int_{\partial\mathcal{M}^5}a_1\smile A_3}\\
		=&\,	Z_{\text{PO}(8k)_{\begin{smallmatrix}-&-&-\\-&-& \end{smallmatrix}}}(-\frac{1}{\tau},B_2^{(1)},B_2^{(2)},A_3)\, .
	\end{split}
\end{align}

We learn that PO$(8k)$ gauge theory has an additional discrete theta angle given by $i\pi\int_{\mathcal{M}^4}(w_1)^2\smile w_2^{(v)}$. A cross-check using the S-duality for Spin$(8k)$ can be performed as before and agrees with this result. Finally, a $T$ transformation takes this boundary condition to 
\begin{align}
	\begin{aligned}
		c_{D1}-b_{F1}\Bigr|_{\substack{\partial\mathcal{M}^5}}=B_2^{(2)} \qquad&&b_{F1}+b_{NS5}+c_{D5}-(a_1)^2\Bigr|_{\substack{\partial\mathcal{M}^5}}=B_2^{(1)}\qquad&&a_3\Bigr|_{\substack{\partial\mathcal{M}^5}}=A_3
	\end{aligned}
\end{align}
which gives us the PO$(8k)_{\begin{smallmatrix}+&-&-\\-&+& \end{smallmatrix}}$ theory.
There is one final duality orbit we need to describe. Consider the boundary condition
\begin{align}
	\begin{aligned}
		b_{F1}\Bigr|_{\substack{\partial\mathcal{M}^5}}=B_2^{(v)} \qquad&&c_{D5}-(a_1)^2\Bigr|_{\substack{\partial\mathcal{M}^5}}=B_2^{(c)}\qquad&&a_3\Bigr|_{\substack{\partial\mathcal{M}^4}}=A_3\, .
	\end{aligned}
\end{align}
Again, all (higher) linking invariants between these operators vanish. The partition function for this theory is given by
\begin{equation}
	Z(\tau,B_2^{(s)},B_2^{(c)},A_3)=\sum_{a_1}Z_{Spin(8k)}(\tau,B_2^{(s)}-(a_1)^2,B_2^{(c)}-(a_1)^2,a_1)(-1)^{\int_{\mathcal{M}^5}a_1\smile A_3}\, .
\end{equation}
This means that in the absence of backgrounds, we sum over PO$(8k)$-bundles with $w_2^{(c)}=(w_1)^2$. We are not aware of any literature studying bundles of this type or the precise nature of this gauge group. We shall refer to this theory as $\widetilde{\text{Pin}}$ gauge theory. If we turn on a background $B_2^{(c)}$, we see that $\delta B_2^{(c)}=\delta (w_2^{(c)}-(w_1)^2)=w_1\smile w_2^{(v)}=w_1\smile(B_2^{(s)}+B_2^{(c)})$. We see that if we turn on a nontrivial background $B_2^{(v)}=B_2^{(s)}+B_2^{(c)}$, then $B_2^{(c)}$ no longer makes sense as a classical background field because $w_1$ is dynamical. This is characteristic of non-invertible symmetries. While the appearance of this theory is somewhat surprising, the remainder of the duality orbit is as expected. The $S$ transformation takes us to the boundary condition
\begin{align}
	\begin{aligned}
		c_{D1}\Bigr|_{\substack{\partial\mathcal{M}^5}}=B_2^{(1)} \qquad&&b_{NS5}-(a_1)^2\Bigr|_{\substack{\partial\mathcal{M}^5}}=B_2^{(2)}\qquad&&a_3\Bigr|_{\substack{\partial\mathcal{M}^4}}=A_3
	\end{aligned}
\end{align}
corresponding to the global form PO$(8k)_{\begin{smallmatrix}+&+&-\\+&+& \end{smallmatrix} }$. From here, a $T$ transformation takes us to
\begin{align}
	\begin{aligned}
		c_{D1}+b_{F1}\Bigr|_{\substack{\partial\mathcal{M}^5}}=B_2^{(1)} \qquad&&b_{NS5}+c_{D5}-(a_1)^2\Bigr|_{\substack{\partial\mathcal{M}^5}}=B_2^{(2)}\qquad&&a_3\Bigr|_{\substack{\partial\mathcal{M}^4}}=A_3
	\end{aligned}
\end{align}
corresponding to the global form PO$(8k)_{\begin{smallmatrix}-&+&-\\+&-& \end{smallmatrix} }$. The resulting duality orbits are summarised in figure \ref{figure:so8k2}.
\subsection*{Absence of additional discrete theta angles}
As was mentioned earlier, the $a_1\smile b_{F1}\smile c_{D1}$ term in the symmetry TFT leads to nontrivial triple-linking between operators in the bulk and therefore obstructs certain gapped boundary conditions. For example, we cannot gauge the non-invertible 0-form symmetry of Ss$(8k)$ and Sc$(8k)$ theories. More interestingly, some of the discrete theta angles of PSO$(8k)$ are not present in the PO$(8k)$ case. Since $\delta w_2^{(c)}=w_2^{(v)}\smile w_1$, $w_2^{(c)}$ is not generally a cocycle here. Therefore, the putative $\frac{1}{2}\mathcal{P}(w_2^{(c)})$ terms become ill-defined. For the same reason, there is no $\frac{1}{2}\mathcal{P}(w_2^{(s)})$ discrete theta angle. However since $w_2^{(v)}$ is always closed, we can still define a discrete theta angle for it. This exactly matches what we found by considering the boundary conditions.

In the preceding section, we have classified all gapped boundary conditions of the schematic form $d_2-(a_1)^2\Bigr|_{\substack{\partial\mathcal{M}^5}}=D_2$, where $d_2$ is any discrete 2-form gauge field in the symmetry TFT. All other boundary conditions of this form which one might attempt to write down are obstructed by a higher link invariant. Consider for example 
\begin{align}
	\begin{aligned}
	b_{F1}-(a_1)^2\Bigr|_{\substack{\partial\mathcal{M}^5}}=B_2^{(1)} \qquad&&c_{D1}+c_{D5}\Bigr|_{\substack{\partial\mathcal{M}^5}}=B_2^{(2)}\qquad&&a_3\Bigr|_{\substack{\partial\mathcal{M}^4}}=A_3
	\end{aligned}
\end{align}
which would serve as a guess for describing Pin$^-(8k)$ gauge theory with  discrete theta angle $\frac{1}{2}\mathcal{P}(w_2^{(c)})$. Of course, in such a theory, $\delta w_2^{(c)}=(w_1)^3$ so we do not expect such a term to be consistent. From the equations of motion of the symmetry TFT, or equivalently the modified Bianchi identity of IIB string theory, we have that $\delta (c_{D1}+c_{D5})=b_{F1}\smile a_1=(b_{F1}-(a_1)^2)\smile a_1+(a_1)^3$. This tells us that we cannot consistently set $(c_{D1}+c_{D5})$ and $(b_{F1}-(a_1)^2)$ equal to background fields on the boundary while allowing $a_1$ to be dynamical. In the language of \cite{Kaidi:2023maf}, the holonomy operators of the given discrete gauge fields have a nontrivial higher linking number obstructing a gapped boundary condition.

All other putative boundary conditions of this form are excluded for similar reasons. This is also consistent with the absence of discrete theta angles of the form $w_2^{(c)}\smile (w_1)^2$ in the PO case. Again, since $w_2^{(c)}$ is not closed in PO$(8k)$ theories, even with background fields turned off, such a term would not be well-defined. From this, we see that boundary conditions of the symmetry TFT can account for all discrete theta angles of the form discussed in \cite{Hsin_2021} while also explaining why they are absent in certain cases.
\subsection*{Other boundary conditions mixing degrees}
Since we allowed for boundary conditions which mix fields of different degrees, there are two more cases we should consider. One might attempt to set $a_3-(a_1)^3\Bigr|_{\substack{\partial\mathcal{M}^5}}=A_3$ or $a_3-d_2\smile a_1\Bigr|_{\substack{\partial\mathcal{M}^5}}=A_3$ for some 2-form gauge-field $d_2$. These boundary conditions do not seem to correspond to any global form of the gauge theory we expect to find. It seems plausible that they are obstructed by self-linking numbers of the corresponding holonomy operators since they involve both $a_1$ and $a_3$ which link nontrivially.

\subsection{$\mathfrak{so}(8k+2)$ theories}
The $\mathfrak{so}(8k+2)$ case is more subtle because of the $\frac{1}{2}\tilde{c}_{D1}\smile\delta\tilde{b}_{F1}$ term in the symmetry TFT. We start with the O$(8k+2)_{++}$ theory. Consider the boundary condition
\begin{align}
	\begin{aligned}
		b_{F1}\Bigr|_{\substack{\partial\mathcal{M}^5}}=B_2^e \qquad&&c_{D1}\Bigr|_{\substack{\partial\mathcal{M}^5}}=B_2^m\qquad&&a_3\Bigr|_{\substack{\partial\mathcal{M}^5}}=A_3\, .
	\end{aligned}
\end{align}
This global form is quite similar to what was found in the previous section; $b_{F1}$ and $c_{D1}$ are closed so we still have a $\mathbb{Z}_2^e\times\mathbb{Z}_2^m$ 1-form symmetry on the boundary.
 The anomaly receives an additional contribution given by $\frac{1}{2}\int_{\mathcal{M}^5}B_2^m\smile\delta \tilde{B}_2^e=\int_{\mathcal{M}^5}B_2^m\smile\beta(B_2^e)$ where $\beta$ is the Bockstein homomorphism associated to the short exact sequence $\mathbb{Z}_2\rightarrow\mathbb{Z}_4\rightarrow\mathbb{Z}_2$. This is consistent with \cite{Hsin_2021}. 
 
 We again attempt to describe the O$(8k+2)_{+-}$ theory with discrete theta angle $\int_{\mathcal{M}^5}w_2^{c}\smile (w_1)^2$ by a boundary condition 
 \begin{align}
 	\begin{aligned}
 		b_{F1}\Bigr|_{\substack{\partial\mathcal{M}^5}}=B_2^e \qquad&&c_{D1}-(a_1)^2\Bigr|_{\substack{\partial\mathcal{M}^5}}=B_2^m\qquad&&a_3\Bigr|_{\substack{\partial\mathcal{M}^5}}=A_3\, .
 	\end{aligned}
 \end{align}
 We then get an additional contribution to the gauge global anomaly compared to the O$(8k)$ case
 \begin{equation}
 	\int_{\mathcal{M}^5}c_{D15}\smile\frac{1}{2} \tilde{b}_{F1} =
 	\int_{\mathcal{M}^5}(c_{D1}-(a_1)^2)\smile\frac{1}{2} \tilde{b}_{F1}+(a_1)^2\smile\frac{1}{2} \tilde{b}_{F1}\, .
 \end{equation}
 The first term on the right hand side is the global anomaly $\int_{\mathcal{M}^5}B_2^m\smile \beta(B_2^e)$ and the second term gives us the contribution to the gauge-global anomaly  $\int_{\mathcal{M}^5}(a_1)^2\smile \beta(B_2^e)$ described in \cite{Hsin_2021}.
 
 We can again consider the action of S-duality on the boundary condition. In the present case, we end up reproducing the same duality orbit as in the previous section.
 
 The remaining cases are slightly more subtle. To prepare, let us revisit the boundary condition for the Spin$(8k+2)$ theory\cite{Etheredge:2023ler}. In this case, D1 branes wrapping $\mathbb{RP}^{1}\in\mathbb{RP}^5$, D5 fat strings, and F1 strings can end on the boundary. The gauge fields under which they are charged satisfy
 \begin{equation}
 	\delta c_{D5}=b_{F1}\smile a_1+\frac{1}{2}\delta \tilde{b}_{F1}\, .
 \end{equation}
 The first term leads to the 2-group structure of the theory while the second one tells us how the two $\mathbb{Z}_2$ 2-form gauge fields combine into a single $\mathbb{Z}_4$ background field on the boundary. From the action \ref{D5action} we see that the defect worldvolume couplings include $c_{D5}$ and $\frac{1}{2}\tilde{c}_{D1}$. So for a D5 to end on the boundary, it is not enough to give Dirichlet boundary conditions to $c_{D5}$. Rather, we need to set
  \begin{align}
 	\begin{aligned}
 		2\tilde{c}_{D5}-\tilde{b}_{F1}\Bigr|_{\substack{\partial\mathcal{M}^5}}=B_2 \qquad&&a_1\Bigr|_{\substack{\partial\mathcal{M}^5}}=A_1
 	\end{aligned}
 \end{align}
 where  $B_2$ is $\mathbb{Z}_4$-valued. The 2-group structure in this case is given by 
 \begin{equation}
 	\delta B_2 = 2 B_2\smile A_1\, .
 \end{equation}
 One must now consider the action of S-duality not only on the discrete gauge fields but also on their $\mathbb{Z}_4$ uplifts, i.e. $\tilde{b}_{F1}\xrightarrow{\text{S}}-\tilde{c}_{D1}$. Note that unlike in our considerations so far, the signs involved actually matter. 
 
 The action of the $S$ generator takes this boundary condition to 
  \begin{align}
 	\begin{aligned}
 		2\tilde{b}_{NS5}+\tilde{c}_{D1}\Bigr|_{\substack{\partial\mathcal{M}^5}}=B_2 \qquad&&a_1\Bigr|_{\substack{\partial\mathcal{M}^5}}=A_1\, .
 	\end{aligned}
 \end{align}
 A subsequent action of $T$ takes us to 
  \begin{align}
 	\begin{aligned}
 		2\tilde{b}_{NS5}+2c_{D5}+\tilde{c}_{D1}-\tilde{b}_{F1}\Bigr|_{\substack{\partial\mathcal{M}^5}}=B_2 \qquad&&a_1\Bigr|_{\substack{\partial\mathcal{M}^5}}=A_1
 	\end{aligned}
 \end{align}
 and from here, $S$ results in 
   \begin{align}
 	\begin{aligned}
 		2\tilde{b}_{NS5}+2c_{D5}+\tilde{c}_{D1}+\tilde{b}_{F1}\Bigr|_{\substack{\partial\mathcal{M}^5}}=B_2 \qquad&&a_1\Bigr|_{\substack{\partial\mathcal{M}^5}}=A_1\, .
 	\end{aligned}
 \end{align}
 Note that for the final boundary condition, a bound state of an NS5 and a $\overline{\text{D5}}$ can end on the boundary while in the previous case, a bound state of a D5 and an NS5 can end. So we are able to reproduce the duality orbit.
 
 We are again interested in the theories we obtain from gauging the 0-form symmetry, i.e. giving Dirichlet boundary conditions to $a_3$. The boundary condition
  \begin{align}
 	\begin{aligned}
 		2\tilde{c}_{D5}-\tilde{b}_{F1}\Bigr|_{\substack{\partial\mathcal{M}^5}}=B_2 \qquad&&a_3\Bigr|_{\substack{\partial\mathcal{M}^5}}=A_3
 	\end{aligned}
 \end{align}
 corresponds to the Pin$^+(8k+2)$ theory as expected. The backgrounds satisfy
 \begin{equation}
 	\delta B_2 =  2B_2\smile a_1\, .
 \end{equation}
 Since $a_1$ does not have Dirichlet boundary conditions, this means we cannot consistently turn on a background $B_2$ and as in the previous examples this indicates the presence of a non-invertible symmetry. However interestingly, as long as $B_2$ is an even cochain, we have $\delta B_2=0$ so the non-invertible symmetry has an invertible $\mathbb{Z}_2$ sub-symmetry. We may proceed as before and derive the duality orbit for this theory. It again mirrors the known duality orbit for the theories with connected gauge group. In particular, the boundary condition for the O$(8k+2)_{-+}$ theory is 
   \begin{align}
  	\begin{aligned}
  		2\tilde{c}_{D5}+2\tilde{c}_{D1}-\tilde{b}_{F1}\Bigr|_{\substack{\partial\mathcal{M}^5}}=B_2 \qquad&&a_3\Bigr|_{\substack{\partial\mathcal{M}^5}}=A_3\, .
  	\end{aligned}
  \end{align}
 As shown in \cite{Hsin_2021}, its gauge-global anomaly receives an additional contribution given by $\pi\int_{\mathcal{M}^5}w_2^{(c)}\smile\beta(w_2^{(v)})$. This is precisely given by the term $\pi\int_{\mathcal{M}^5}c_{D1}\smile\delta\tilde{b}_{F1}$ in the symmetry TFT. For a summary of these results see figure \ref{figure:so8kplus21}.
 
 Finally, we may consider boundary conditions of the form 
 
 \begin{align}
 	\begin{aligned}
 		2\tilde{c}_{D5}-2(\tilde{a}_1)^2-\tilde{b}_{F1}\Bigr|_{\substack{\partial\mathcal{M}^5}}=B_2 \qquad&&a_3\Bigr|_{\substack{\partial\mathcal{M}^5}}=A_3\, .
 	\end{aligned}
 \end{align}
 
 This corresponds to the $\widetilde{\text{Pin}}(8k+2)$ theory. We again obtain a duality orbit mirroring the already known ones, shown in figure \ref{figure:so8kplus22}.

\section{Fusion rules}
In the previous section, we made statements about which duality orbits were expected to have non-invertible symmetries. In this section, we shall compute the relevant fusion rules to confirm these expectations. Note that the operator arising from a D3 brane wrapping $\mathbb{RP}^3\subset\mathbb{RP}^5$ is not stacked with a nontrivial TQFT so the fusion of the corresponding symmetry defects is that of $\mathbb{Z}_2$ for all boundary conditions. Therefore, we shall focus only on the fusion rules of the 1-form symmetry.
\subsection{$\mathfrak{so}(8k)$ theories}
We start with the O$(4k)_{++}$-theory. A possible choice of symmetry operators for the 1-form part is D5 and NS5 branes wrapping $\mathbb{RP}^4$$\subset\mathbb{RP}^5$. We might choose different sets of branes as our symmetry operators, such as D1/D5 and F1/NS5 bound states. In cases in which the 1-form symmetry is not anomalous (not including mixed anomalies with symmetries of different degrees), we may choose a commuting set of branes. These data, together with the boundary conditions for the bulk gauge fields define a polarization pair \cite{Lawrie:2023tdz}. As explained there, changing the choice of symmetry operators, corresponds to stacking with an SPT-phase for the symmetry. We will not classify all such choices in this paper and instead will be content with calculating fusion rules for one possible set of symmetry generators in each case.

For $N=8k$, the TQFT on the D5 worldvolume simplifies to

 \begin{equation}\label{D58k}
	D5(\Sigma^2)=\mathcal{D}\gamma_1\mathcal{D}\phi_0\exp(i \pi \int_{\Sigma^2}c_{D5}+\gamma_1\smile\delta\phi_0+\phi_0\smile b_{F1}+\gamma_1\smile a_1)\, .
\end{equation}
We may calculate the fusion of two D5 branes pushed to the boundary similarly to \cite{Kaidi:2021xfk}\cite{Bhardwaj_2023}. Since $b_{F1}\Bigr|_{\substack{\partial\mathcal{M}^5}}=B_2^e$, the $\phi_0\smile b_{F1}$ term will not play a role here; it will only change the fusion rules by an overall constant and throughout this section, we shall not attempt to determine the correct normalisation for the fusion rules. One arrives at

\begin{equation}
   D5(\Sigma^2)\times D5(\Sigma^2)\propto\sum_{\Gamma\in H_1(\Sigma^2;\mathbb{Z}_2)}\exp(i \pi \int_{\Gamma}a_1)
\end{equation}
and similarly for NS5-branes. This is consistent with our expectation for a 3-group symmetry.

Consider now the O$(8k)_{-+}$-theory. A possible set of symmetry operators is again given by NS5 fat strings and D1 branes. The fusion of D5 branes is unchanged since $b_{F1}$ still has a Dirichlet boundary condition. However, now the combination $c_{D1}-c_{D5}$ has a Dirichlet boundary condition. We rewrite the topological sector of the NS5 worldvolume theory as 
\begin{equation}
	NS5(\Sigma^2)=\mathcal{D}\gamma_1\mathcal{D}\phi_0\exp(i \pi \int_{\Sigma^2}b_{NS5}+\gamma_1\smile\delta\phi_0+\phi_0\smile c_{D1}+\gamma_1\smile a_1)\, .
\end{equation}
One can check that this TQFT cancels the gauge anomaly given by $\pi i \int_{\Sigma^2}w_2^{c}\smile w_1$ on the defect worldvolume calculated in \cite{Hsin_2021} by noting that on the boundary, $c_{D5}$ is identified with $w_2^{c}$ and differs from $c_{D1}$ by a c-number. Equivalently, this cancels the gauge-global anomaly, which previously prevented us from turning on a background for $B_2^e$. Assuming $\Sigma^2$ to be connected, this results in the fusion rules:
\begin{equation}
	NS5(\Sigma^2)\times NS5(\Sigma^2)\propto (1+D1(\Sigma^2))\sum_{\Gamma\in H_1(\Sigma^2;\mathbb{Z}_2)}\exp(i \pi \int_{\Gamma}a_1)\, .
\end{equation}
One might ask about the possibility of making a different choice of symmetry operators such as NS5 and D5. Recall the D5 worldvolume theory
\begin{equation}
	D5(\Sigma^2)=\mathcal{D}\gamma_1\mathcal{D}\phi_0\exp(i \pi \int_{\Sigma^2}c_{D5}+\gamma_1\smile\delta\phi_0+\phi_0\smile b_{F1}+\gamma_1\smile a_1)\, .
\end{equation}
Here,  $b_{F1}$ is the background for the non-invertible electric 1-form symmetry. As argued in \cite{Hsin_2021}, it is trivial in cohomology so we take it to be zero as a cochain. We may now perform the path integrals on the worldvolume, resulting in 
\begin{equation}
	D5(\Sigma^2)\propto \sum_{\Gamma\in H_1(\Sigma^2;\mathbb{Z}_2)}\exp(i \pi \int_{\Sigma^2}c_{D5}+\int_{\Gamma}a_1)\, .
\end{equation} 
Recalling that $c_{D1}$ and $c_{D5}$ differ by a c-number on the boundary, we conclude that a D5 fat string can be obtained from a D1 by gauging the 3-form symmetry on the D1 worldvolume. It has been discussed in the literature how different choices of branes symmetry generators differ by SPT phases and how these data are encoded by polarisation pairs \cite{Lawrie:2023tdz}. Our observation suggests that this story is modified since the D1 appears to be the simple defect, from which the D5 fat string is obtained by condensation \cite{Choi:2022zal}.

The fusion rules for the Pin$^+(8k)$ theory have already been computed in \cite{Bhardwaj_2023}. A possible set of symmetry generators is D1, NS5. The D1 obey $\mathbb{Z}_2$ fusion rules. The D5 worldvolume theory (\ref{D58k}) matches the TQFT stacked on the topological defect proposed in \cite{Bhardwaj_2023} and therefore, the fusion rules match as well.

Let us now turn to the theories obtained from non-simple boundary conditions. A possible set of symmetry operators for the O$(8k)_{+-}$-theory is given by D5 and NS5 fat strings. The fusion of two D5 branes is the same as in the O$(8k)_{++}$ case. If we set $B_2^m=0$, the worldvolume theory of an NS5 pushed to the boundary becomes
\begin{equation}
		NS5(\Sigma^2)=\mathcal{D}\gamma_1\mathcal{D}\phi_0\exp(i\pi \int_{\Sigma^2}b_{NS5}+\gamma_1\smile\delta\phi_0+\phi_0\smile (a_1)^2+\gamma_1\smile a_1)\, .
\end{equation}
As before, turning on the background will only change this discussion by a constant factor. Again, the gauge-global anomaly of this theory may be viewed as a gauge anomaly on the defect worldvolume given by $\pi i \int_{\Sigma^3}(w_1)^3$ \cite{Hsin_2021}. We see that the above action cancels this anomaly. The defects fuse as
\begin{equation}
	NS5(\Sigma^2)\times NS5(\Sigma^2)=(1+\exp(i\pi \int_{\Sigma^2}(a_1)^2))\sum_{\Gamma\in H_1(\Sigma^2;\mathbb{Z}_2)}\exp(i\pi \int_{\Gamma}a_1) \, .
\end{equation}
Interestingly, the first factor on the right hand side is only nontrivial when $\Sigma^2$ is unorientable.

The O$(8k)_{--}$ theory is a combination of the previous two cases. We take NS5 and D1 branes as our symmetry generators. The fusion of two D1 branes is invertible as always. The NS5 worldvolume theory is
\begin{equation}
	NS5(\Sigma^2)=\mathcal{D}\gamma_1\mathcal{D}\phi_0\exp(i\pi \int_{\Sigma^2}b_{NS5}+\gamma_1\smile\delta\phi_0+\phi_0\smile c_{D1}+\gamma_1\smile a_1)\, .
\end{equation}
Hence the fusion is
\begin{equation}
	NS5(\Sigma^2)\times NS5(\Sigma^2)=(1+D1(\Sigma^2))\sum_{\Gamma\in H_1(\Sigma^2;\mathbb{Z}_2)}\exp(i\pi \int_{\Gamma}a_1)\, .
\end{equation}
Finally, for the $\widetilde{\text{Pin}}(8k)$ theory, we may choose D1 and NS5 branes. These operators turn out to have the same fusion rules as in the Pin$^+(8k)$-theory. We have thus calculated the relevant fusion rules for a representative of each duality orbit in the $\mathfrak{so}(8k)$-case with disconnected gauge group.
\subsection{$\mathfrak{so}(8k+2)$ theories}
In this case, the duality orbits are bigger so there are fewer examples we need to study. The fusion rules in the O$(8k+2)_{++}$ case are the same as in the O$(8k)_{++}$ one. Next is the duality orbit including Pin$^+(8k+2)$. The fusion rules in this case were computed in \cite{Bhardwaj_2023} and since the action \ref{D5action} agrees with the action proposed in that paper, so do the fusion rules. 

In the O$(8k+2)_{+-}$ theory, we turn off both backgrounds for the 1-form symmetry. The NS5 worldvolume theory becomes
\begin{equation}
	NS5(\Sigma^2)=\mathcal{D}\gamma_1\mathcal{D}\phi_0\exp(i\pi \int_{\Sigma^2}b_{NS5}+\gamma_1\smile\delta\phi_0+\phi_0\smile (a_1)^2+\gamma_1\smile a_1+\frac{1}{2}(\delta\tilde{\gamma}_1-(\tilde{a}_1)^2))
\end{equation}
leading to the fusion rules
\begin{equation}
	NS5(\Sigma^2)\times NS5(\Sigma^2)\propto(1+\exp(i\pi \int_{\Sigma^2}(a_1)^2))\sum_{\Gamma\in H_1(\Sigma^2;\mathbb{Z}_2)}\exp(i\pi Q(\Gamma))\exp(\pi i \int_{\Gamma}a_1)
\end{equation}
where $Q(\Gamma)=\int_{\Sigma^2}\text{Bock}(\text{PD}(\Gamma))$ as in \cite{Bhardwaj_2023}. The fusion of D5 generators is the same as in the O$(8k+2)_{++}$ theory.

We are left with one more duality orbit, for which we choose the O$(8k+2)_{--}$ theory as our representative. Taking the generators to be NS5 and D1, yields the same fusion rules as Pin$^+(8k+2)$. 

\acknowledgments

The author thanks Ben Heidenreich, Muldrow Etheredge, and Sebastian Rauch for helpful discussions and collaboration on related projects, I\~naki Garc\'ia Etxebarria for initial discussions that led to this paper, and Muldrow Etheredge and I\~naki Garc\'ia Etxebarria for their comments on earlier drafts of this paper. FBC is supported by an STFC training grant, project reference 2713393.
\appendix
\section{K\"unneth theorem and torsion}\label{sec:Kuenneth}
The (algebraic) Künneth theorem states that for cochain complexes $C$, $C'$
\begin{equation}
	\label{Kunneth}
	0\rightarrow\bigoplus_{i+j=n}H^i(C)\otimes H^j(C')\rightarrow H^n(C\otimes  C')\rightarrow\bigoplus_{i+j=n+1}\text{Tor}_1(H^i(C), H^j(C'))\rightarrow 0
\end{equation}
splits non-canonically. We are restricting ourselves to cohomology with integer coefficients for the time being but shall make some comments about a generalisation to cohomology with local coefficients at the end of this subsection. The map $\bigoplus_{i+j=n}H^i(C)\otimes H^j(C')\rightarrow H^n(C\otimes C')$ has been discussed in the physics literature via projection maps and cup products \cite{Apruzzi:2021nmk} (see e.g. \cite{Davis} lemma 3.10 for a full mathematical treatment). However, contributions from the final term of the sequence are more subtle and have so far been ignored. We know that $\text{Tor}(H^i(C)\otimes H^j(C'))$ is nontrivial only if both $H^i(C)$ and $H^j(C')$ have torsion elements. For the sake of definiteness, let $n=3$, $H^3(C)=\mathbb{Z}^l\oplus\mathbb{Z}_{2}$ with $[s_3]$ generating the  $\mathbb{Z}_{2}$ subgroup and $H^1(C')=\mathbb{Z}_2$ generated by $[t_1]$ and assume all other cohomology groups are torsion-free. Then $\text{Tor}(H^i(C),H^i(C'))=\mathbb{Z}_2$. 

We now want to understand what an element in $H^n(C\otimes C')$ that gets mapped to the generator of this group looks like. Following the proof in \cite{Hatcher}, we have a SES $0\rightarrow Z\rightarrow C \xrightarrow{\partial} B \rightarrow 0$ of chain complex where the complexes of cycles and boundaries each have trivial differential maps. Tensoring this with $C'$ gives us another SES of chain complexes $0\rightarrow Z\otimes C'\rightarrow C\otimes C' \xrightarrow{\text{d}\otimes \mathrm{id}} B\otimes C' \rightarrow 0$. A short exact sequence of cochain complexes leads to a long exact sequence in cohomology. Diagram-chasing arguments show that the connecting homomorphism is the inclusion map $B^n\rightarrow Z^n$ tensored with the identity on $C'$. Using also the fact that the differential maps on $B$ and $Z$ are trivial, the LES becomes: 
$$
...\xrightarrow{i_n}\bigoplus_i(Z^i \otimes H^{n-i}(C'))\rightarrow H^n(C\otimes C')\rightarrow\bigoplus_i
(B^i\otimes H^{n+1-i}(C'))\xrightarrow{i_{n+1}}...$$
From this, we get a short exact sequence for each $n$:
$$
0\rightarrow\text{Coker}i_n\rightarrow H^n(C\otimes C')\rightarrow \text{Ker}i_{n+1}\rightarrow 0
$$
and the first and third nontrivial terms can be shown to be isomorphic to the corresponding terms in the Künneth short exact sequence. Our task is now to find an element in $\text{Ker}i_{n+1}$ for $n=3$. Since Tor$(A,B)$ is nontrivial only if both $A$ and $B$ have torsion elements, we know that this should involve $s_3$ and $t_1$. Since $2s_3=dp_2\in B^3$, a natural guess is $dp_2\otimes[t_1]$. Indeed, its image $i_{n+1}(dp_2\otimes[t_1])=2s_3\otimes [t_1]=s_3\otimes 2[t_1]=0$ (note that this does not imply that $dp_2\otimes[t_1]=0\in B^3\otimes H^1(C')$ because $s_3\notin B^3$). Its pre-image in $H^3(C\otimes C')$ consists of elements of the form $[p_2\otimes t_1+...]$. Any two elements will differ by the image of an element in $\text{Coker}i_n$ by exactness so it suffices to construct one element in the pre-image. $p_1\otimes t_1$ is not closed as $\delta(p_1\otimes t_1)=2s_3\otimes t_1=s_3\otimes 2t_1=s_3\otimes \delta h_0$ for some cochain $h_0$ but the combination $p_2\otimes t_1 + s_3\otimes h_0$ is. The cohomology class $[p_2\otimes t_1 + s_3\otimes h_0]$ then gets mapped to $dp_2\otimes [t_1]\in \text{Ker}i_{n+1}$. This generalises straightforwardly to the case where $[s_3]$ is $2q$-torsion for $q\in\mathbb{N}$ since Tor$(\mathbb{Z}_2,\mathbb{Z}_{2q})=\mathbb{Z}_2$: simply replace $[s_3]$ by $q[s_3]$, which is 2-torsion.

We now want to apply this discussion to the cohomology of product spaces. As shown in \cite{spanier1989algebraic}, there is a chain homotopy between $C^*(X\times Y)$ and $C^*(X)\otimes C^*(Y)$ given by the cohomology cross product, or equivalently by the cup product of the pullbacks under the projection maps $X\times Y\rightarrow X$ and $X\times Y\rightarrow Y$. This allows us to write down representative cocycles in the form familiar from \cite{Apruzzi:2021nmk} and as in that paper, we will not explicitly write down the pullbacks.

Since the  geometry we are considering includes an orientifold, we need to consider cohomology with local coefficients. The K\"unneth theorem in cohomology was generalised to the case of local coefficients in \cite{greenblatt2006homology} for spaces which admit a finite cell-complex structure so we expect the discussion to apply in this case as well.
\paragraph{Data access statement.} There is no additional research
data associated with this work.

\appendix

\bibliographystyle{JHEP}
\bibliography{refs}

\providecommand{\href}[2]{#2}\begingroup\raggedright\begin{thebibliography}{10}

\bibitem{Witten_1998}
E.~Witten, {\it Baryons and branes in anti de sitter space},  {\em Journal of
  High Energy Physics} {\bf 1998} (July, 1998) 006–006.

\bibitem{Gaiotto:2010be}
D.~Gaiotto, G.~W. Moore, and A.~Neitzke, {\it {Framed BPS States}},  {\em Adv.
  Theor. Math. Phys.} {\bf 17} (2013), no.~2 241--397,
  [\href{http://arxiv.org/abs/1006.0146}{{\tt arXiv:1006.0146}}].

\bibitem{Aharony:2013hda}
O.~Aharony, N.~Seiberg, and Y.~Tachikawa, {\it {Reading between the lines of
  four-dimensional gauge theories}},  {\em JHEP} {\bf 08} (2013) 115,
  [\href{http://arxiv.org/abs/1305.0318}{{\tt arXiv:1305.0318}}].

\bibitem{Bergman_2022}
O.~Bergman and S.~Hirano, {\it The holography of duality in $ \mathcal{N} $ = 4
  super-yang-mills theory},  {\em Journal of High Energy Physics} {\bf 2022}
  (Nov., 2022).

\bibitem{Etheredge:2023ler}
M.~Etheredge, I.~García~Etxebarria, B.~Heidenreich, and S.~Rauch, {\it {Branes
  and symmetries for $ \mathcal{N} $ = 3 S-folds}},  {\em JHEP} {\bf 09} (2023)
  005, [\href{http://arxiv.org/abs/2302.14068}{{\tt arXiv:2302.14068}}].

\bibitem{Hsin_2021}
P.-S. Hsin and H.~T. Lam, {\it Discrete theta angles, symmetries and
  anomalies},  {\em SciPost Physics} {\bf 10} (Feb., 2021).

\bibitem{Gaiotto:2020iye}
D.~Gaiotto and J.~Kulp, {\it {Orbifold groupoids}},  {\em JHEP} {\bf 02} (2021)
  132, [\href{http://arxiv.org/abs/2008.05960}{{\tt arXiv:2008.05960}}].

\bibitem{Apruzzi:2021nmk}
F.~Apruzzi, F.~Bonetti, I.~García~Etxebarria, S.~S. Hosseini, and
  S.~Schafer-Nameki, {\it {Symmetry TFTs from String Theory}},  {\em Commun.
  Math. Phys.} {\bf 402} (2023), no.~1 895--949,
  [\href{http://arxiv.org/abs/2112.02092}{{\tt arXiv:2112.02092}}].

\bibitem{Freed:2022qnc}
D.~S. Freed, G.~W. Moore, and C.~Teleman, {\it {Topological symmetry in quantum
  field theory}},  \href{http://arxiv.org/abs/2209.07471}{{\tt
  arXiv:2209.07471}}.

\bibitem{GarciaEtxebarria:2022vzq}
I.~García~Etxebarria, {\it {Branes and Non-Invertible Symmetries}},  {\em
  Fortsch. Phys.} {\bf 70} (2022), no.~11 2200154,
  [\href{http://arxiv.org/abs/2208.07508}{{\tt arXiv:2208.07508}}].

\bibitem{Apruzzi:2022rei}
F.~Apruzzi, I.~Bah, F.~Bonetti, and S.~Schafer-Nameki, {\it {Noninvertible
  Symmetries from Holography and Branes}},  {\em Phys. Rev. Lett.} {\bf 130}
  (2023), no.~12 121601, [\href{http://arxiv.org/abs/2208.07373}{{\tt
  arXiv:2208.07373}}].

\bibitem{Heckman:2022muc}
J.~J. Heckman, M.~H\"ubner, E.~Torres, and H.~Y. Zhang, {\it {The Branes Behind
  Generalized Symmetry Operators}},  {\em Fortsch. Phys.} {\bf 71} (2023),
  no.~1 2200180, [\href{http://arxiv.org/abs/2209.03343}{{\tt
  arXiv:2209.03343}}].

\bibitem{Gaiotto:2014kfa}
D.~Gaiotto, A.~Kapustin, N.~Seiberg, and B.~Willett, {\it {Generalized Global
  Symmetries}},  {\em JHEP} {\bf 02} (2015) 172,
  [\href{http://arxiv.org/abs/1412.5148}{{\tt arXiv:1412.5148}}].

\bibitem{Sharpe_2015}
E.~Sharpe, {\it Notes on generalized global symmetries in qft},  {\em
  Fortschritte der Physik} {\bf 63} (Sept., 2015) 659–682.

\bibitem{Tachikawa:2017gyf}
Y.~Tachikawa, {\it {On gauging finite subgroups}},  {\em SciPost Phys.} {\bf 8}
  (2020), no.~1 015, [\href{http://arxiv.org/abs/1712.09542}{{\tt
  arXiv:1712.09542}}].

\bibitem{Benini:2018reh}
F.~Benini, C.~C\'ordova, and P.-S. Hsin, {\it {On 2-Group Global Symmetries and
  their Anomalies}},  {\em JHEP} {\bf 03} (2019) 118,
  [\href{http://arxiv.org/abs/1803.09336}{{\tt arXiv:1803.09336}}].

\bibitem{Bhardwaj:2021wif}
L.~Bhardwaj, {\it {2-Group Symmetries in Class S}},
  \href{http://arxiv.org/abs/2107.06816}{{\tt arXiv:2107.06816}}.

\bibitem{DelZotto:2022joo}
M.~Del~Zotto, I.~García~Etxebarria, and S.~Schäfer-Nameki, {\it {2-Group
  Symmetries and M-Theory}},  \href{http://arxiv.org/abs/2203.10097}{{\tt
  arXiv:2203.10097}}.

\bibitem{Apruzzi:2021vcu}
F.~Apruzzi, L.~Bhardwaj, J.~Oh, and S.~Schafer-Nameki, {\it {The Global Form of
  Flavor Symmetries and 2-Group Symmetries in 5d SCFTs}},
  \href{http://arxiv.org/abs/2105.08724}{{\tt arXiv:2105.08724}}.

\bibitem{Heidenreich:2021xpr}
B.~Heidenreich, J.~McNamara, M.~Montero, M.~Reece, T.~Rudelius, and
  I.~Valenzuela, {\it {Non-invertible global symmetries and completeness of the
  spectrum}},  {\em JHEP} {\bf 09} (2021) 203,
  [\href{http://arxiv.org/abs/2104.07036}{{\tt arXiv:2104.07036}}].

\bibitem{Kaidi:2021xfk}
J.~Kaidi, K.~Ohmori, and Y.~Zheng, {\it {Kramers-Wannier-like Duality Defects
  in (3+1)D Gauge Theories}},  {\em Phys. Rev. Lett.} {\bf 128} (2022), no.~11
  111601, [\href{http://arxiv.org/abs/2111.01141}{{\tt arXiv:2111.01141}}].

\bibitem{Choi:2021kmx}
Y.~Choi, C.~Cordova, P.-S. Hsin, H.~T. Lam, and S.-H. Shao, {\it {Noninvertible
  duality defects in 3+1 dimensions}},  {\em Phys. Rev. D} {\bf 105} (2022),
  no.~12 125016, [\href{http://arxiv.org/abs/2111.01139}{{\tt
  arXiv:2111.01139}}].

\bibitem{Cordova_2018}
C.~Cordova, P.-S. Hsin, and N.~Seiberg, {\it Global symmetries, counterterms,
  and duality in chern-simons matter theories with orthogonal gauge groups},
  {\em SciPost Physics} {\bf 4} (Apr., 2018).

\bibitem{Bhardwaj_2023}
L.~Bhardwaj, L.~E. Bottini, S.~Schäfer-Nameki, and A.~Tiwari, {\it
  Non-invertible symmetry webs},  {\em SciPost Physics} {\bf 15} (Oct., 2023).

\bibitem{Bartsch_2024}
T.~Bartsch, M.~Bullimore, A.~E.~V. Ferrari, and J.~Pearson, {\it Non-invertible
  symmetries and higher representation theory i},  {\em SciPost Physics} {\bf
  17} (July, 2024).

\bibitem{Bartsch_2024i}
T.~Bartsch, M.~Bullimore, A.~E.~V. Ferrari, and J.~Pearson, {\it Non-invertible
  symmetries and higher representation theory ii},  {\em SciPost Physics} {\bf
  17} (Aug., 2024).

\bibitem{Bergman:2024its}
O.~Bergman and F.~Mignosa, {\it {String theory and the SymTFT of 3d
  orthosymplectic Chern-Simons theory}},
  \href{http://arxiv.org/abs/2412.00184}{{\tt arXiv:2412.00184}}.

\bibitem{Bonetti:2024etn}
F.~Bonetti, M.~Del~Zotto, and R.~Minasian, {\it {SymTFTs and Non-Invertible
  Symmetries of 6d (2,0) SCFTs of Type $D$ from M-theory}},
  \href{http://arxiv.org/abs/2412.07842}{{\tt arXiv:2412.07842}}.

\bibitem{GarciaEtxebarria:2024fuk}
I.~García~Etxebarria and S.~S. Hosseini, {\it {Some aspects of symmetry
  descent}},  \href{http://arxiv.org/abs/2404.16028}{{\tt arXiv:2404.16028}}.

\bibitem{Moore:1999gb}
G.~W. Moore and E.~Witten, {\it {Selfduality, Ramond-Ramond fields, and K
  theory}},  {\em JHEP} {\bf 05} (2000) 032,
  [\href{http://arxiv.org/abs/hep-th/9912279}{{\tt hep-th/9912279}}].

\bibitem{Freed:2000ta}
D.~S. Freed, {\it {Dirac charge quantization and generalized differential
  cohomology}},  \href{http://arxiv.org/abs/hep-th/0011220}{{\tt
  hep-th/0011220}}.

\bibitem{diaconescu2004e8gaugetheoryderivation}
D.-E. Diaconescu, G.~Moore, and E.~Witten, {\it E8 gauge theory, and a
  derivation of k-theory from m-theory},  2004.

\bibitem{Evslin:2006cj}
J.~Evslin, {\it {What does(n't) K-theory classify?}},
  \href{http://arxiv.org/abs/hep-th/0610328}{{\tt hep-th/0610328}}.

\bibitem{10.1007/BFb0075216}
J.~Cheeger and J.~Simons, {\it Differential characters and geometric
  invariants},  in {\em Geometry and Topology}, (Berlin, Heidelberg),
  pp.~50--80, Springer Berlin Heidelberg, 1985.

\bibitem{bär2014differential}
C.~B{\"a}r and C.~Becker, {\em Differential Characters}.
\newblock Lecture Notes in Mathematics. Springer International Publishing,
  2014.

\bibitem{Freed:2006yc}
D.~S. Freed, G.~W. Moore, and G.~Segal, {\it {Heisenberg Groups and
  Noncommutative Fluxes}},  {\em Annals Phys.} {\bf 322} (2007) 236--285,
  [\href{http://arxiv.org/abs/hep-th/0605200}{{\tt hep-th/0605200}}].

\bibitem{Hopkins:2002rd}
M.~J. Hopkins and I.~M. Singer, {\it {Quadratic functions in geometry,
  topology, and M theory}},  {\em J. Diff. Geom.} {\bf 70} (2005), no.~3
  329--452, [\href{http://arxiv.org/abs/math/0211216}{{\tt math/0211216}}].

\bibitem{Belov:2006jd}
D.~Belov and G.~W. Moore, {\it {Holographic Action for the Self-Dual Field}},
  \href{http://arxiv.org/abs/hep-th/0605038}{{\tt hep-th/0605038}}.

\bibitem{Belov:2006xj}
D.~M. Belov and G.~W. Moore, {\it {Type II Actions from 11-Dimensional
  Chern-Simons Theories}},  \href{http://arxiv.org/abs/hep-th/0611020}{{\tt
  hep-th/0611020}}.

\bibitem{Hsieh:2020jpj}
C.-T. Hsieh, Y.~Tachikawa, and K.~Yonekura, {\it {Anomaly Inflow and p-Form
  Gauge Theories}},  {\em Commun. Math. Phys.} {\bf 391} (2022), no.~2
  495--608, [\href{http://arxiv.org/abs/2003.11550}{{\tt arXiv:2003.11550}}].

\bibitem{Hatcher}
A.~Hatcher, {\em Algebraic Topology}.
\newblock Algebraic Topology. Cambridge University Press, 2002.

\bibitem{Heckman:2022xgu}
J.~J. Heckman, M.~Hubner, E.~Torres, X.~Yu, and H.~Y. Zhang, {\it {Top down
  approach to topological duality defects}},  {\em Phys. Rev. D} {\bf 108}
  (2023), no.~4 046015, [\href{http://arxiv.org/abs/2212.09743}{{\tt
  arXiv:2212.09743}}].

\bibitem{Apruzzi:2023uma}
F.~Apruzzi, F.~Bonetti, D.~S.~W. Gould, and S.~Schafer-Nameki, {\it {Aspects of
  Categorical Symmetries from Branes: SymTFTs and Generalized Charges}},
  \href{http://arxiv.org/abs/2306.16405}{{\tt arXiv:2306.16405}}.

\bibitem{Bah:2023ymy}
I.~Bah, E.~Leung, and T.~Waddleton, {\it {Non-invertible symmetries, brane
  dynamics, and tachyon condensation}},  {\em JHEP} {\bf 01} (2024) 117,
  [\href{http://arxiv.org/abs/2306.15783}{{\tt arXiv:2306.15783}}].

\bibitem{Bhardwaj_2023i}
L.~Bhardwaj, L.~E. Bottini, S.~Schäfer-Nameki, and A.~Tiwari, {\it
  Non-invertible higher-categorical symmetries},  {\em SciPost Physics} {\bf
  14} (Jan., 2023).

\bibitem{bhardwaj2023generalizedchargesiinoninvertible}
L.~Bhardwaj and S.~Schafer-Nameki, {\it Generalized charges, part ii:
  Non-invertible symmetries and the symmetry tft},  2023.

\bibitem{Kapustin_2011}
A.~Kapustin and N.~Saulina, {\it Topological boundary conditions in abelian
  chern–simons theory},  {\em Nuclear Physics B} {\bf 845} (Apr., 2011)
  393–435.

\bibitem{bhardwaj2024gappedphases21dnoninvertible}
L.~Bhardwaj, D.~Pajer, S.~Schafer-Nameki, A.~Tiwari, A.~Warman, and J.~Wu, {\it
  Gapped phases in (2+1)d with non-invertible symmetries: Part i},  2024.

\bibitem{Bullimore:2024khm}
M.~Bullimore and J.~J. Pearson, {\it {Towards All Categorical Symmetries in 2+1
  Dimensions}},  \href{http://arxiv.org/abs/2408.13931}{{\tt
  arXiv:2408.13931}}.

\bibitem{Kaidi:2023maf}
J.~Kaidi, E.~Nardoni, G.~Zafrir, and Y.~Zheng, {\it {Symmetry TFTs and
  anomalies of non-invertible symmetries}},  {\em JHEP} {\bf 10} (2023) 053,
  [\href{http://arxiv.org/abs/2301.07112}{{\tt arXiv:2301.07112}}].

\bibitem{Kaidi_2022}
J.~Kaidi, G.~Zafrir, and Y.~Zheng, {\it Non-invertible symmetries of $
  \mathcal{N} $ = 4 sym and twisted compactification},  {\em Journal of High
  Energy Physics} {\bf 2022} (Aug., 2022).

\bibitem{Zhang:2024oas}
H.~Y. Zhang, {\it {K-theoretic Global Symmetry in String-constructed QFT and
  T-duality}},  \href{http://arxiv.org/abs/2404.16097}{{\tt arXiv:2404.16097}}.

\bibitem{kirby1990pin}
R.~C. Kirby and L.~R. Taylor, {\it Pin structures on low-dimensional
  manifolds},  {\em Geometry of low-dimensional manifolds} {\bf 2} (1990)
  177--242.

\bibitem{Kaidi:2022cpf}
J.~Kaidi, K.~Ohmori, and Y.~Zheng, {\it {Symmetry TFTs for Non-invertible
  Defects}},  {\em Commun. Math. Phys.} {\bf 404} (2023), no.~2 1021--1124,
  [\href{http://arxiv.org/abs/2209.11062}{{\tt arXiv:2209.11062}}].

\bibitem{doi:10.1142/S0218216598000206}
W.~S. MASSEY, {\it Higher order linking numbers},  {\em Journal of Knot Theory
  and Its Ramifications} {\bf 07} (1998), no.~03 393--414,
  [\href{http://arxiv.org/abs/https://doi.org/10.1142/S0218216598000206}{{\tt
  https://doi.org/10.1142/S0218216598000206}}].

\bibitem{Putrov_2017}
P.~Putrov, J.~Wang, and S.-T. Yau, {\it Braiding statistics and link invariants
  of bosonic/fermionic topological quantum matter in 2+1 and 3+1 dimensions},
  {\em Annals of Physics} {\bf 384} (Sept., 2017) 254–287.

\bibitem{Wan:2019oyr}
Z.~Wan, J.~Wang, and Y.~Zheng, {\it {Quantum 4d Yang-Mills Theory and
  Time-Reversal Symmetric 5d Higher-Gauge Topological Field Theory}},  {\em
  Phys. Rev. D} {\bf 100} (2019), no.~8 085012,
  [\href{http://arxiv.org/abs/1904.00994}{{\tt arXiv:1904.00994}}].

\bibitem{Zhang_2021}
Z.-F. Zhang and P.~Ye, {\it Compatible braidings with hopf links, multiloop,
  and borromean rings in $(3+1)$-dimensional spacetime},  {\em Physical Review
  Research} {\bf 3} (May, 2021).

\bibitem{Zhang_2022}
Z.-F. Zhang and P.~Ye, {\it Topological orders, braiding statistics, and
  mixture of two types of twisted bf theories in five dimensions},  {\em
  Journal of High Energy Physics} {\bf 2022} (Apr., 2022).

\bibitem{Del_Zotto_2024}
M.~Del~Zotto, S.~N. Meynet, and R.~Moscrop, {\it Remarks on geometric
  engineering, symmetry tfts and anomalies},  {\em Journal of High Energy
  Physics} {\bf 2024} (July, 2024).

\bibitem{Lawrie:2023tdz}
C.~Lawrie, X.~Yu, and H.~Y. Zhang, {\it {Intermediate Defect Groups,
  Polarization Pairs, and Non-invertible Duality Defects}},
  \href{http://arxiv.org/abs/2306.11783}{{\tt arXiv:2306.11783}}.

\bibitem{Choi:2022zal}
Y.~Choi, C.~Cordova, P.-S. Hsin, H.~T. Lam, and S.-H. Shao, {\it
  {Non-invertible Condensation, Duality, and Triality Defects in 3+1
  Dimensions}},  {\em Commun. Math. Phys.} {\bf 402} (2023), no.~1 489--542,
  [\href{http://arxiv.org/abs/2204.09025}{{\tt arXiv:2204.09025}}].

\bibitem{Davis}
J.~F. Davis, {\em Lecture notes in algebraic topology / James F. Davis, Paul
  Kirk.}
\newblock Graduate studies in mathematics ; v 35. American Mathematical
  Society, 2001.

\bibitem{spanier1989algebraic}
E.~Spanier, {\em Algebraic Topology}.
\newblock McGraw-Hill series in higher mathematics. Springer, 1989.

\bibitem{greenblatt2006homology}
R.~Greenblatt, {\it Homology with local coefficients and characteristic
  classes},  {\em Homology, Homotopy and Applications} {\bf 8} (2006), no.~2
  91--103.

\end{thebibliography}\endgroup

\end{document}